\documentclass[epj,natbib,sort&compress]{svjour}
\usepackage{graphics}
\usepackage{amsmath}
\usepackage{amsfonts}
\usepackage{amssymb}
\usepackage{mathtools}
\usepackage[utf8]{inputenc}
\usepackage[T1]{fontenc}
\usepackage[colorlinks=true]{hyperref}
\usepackage[sort&compress,numbers]{natbib}

\newcommand{\dd}{\mathrm{d}}
\newcommand{\Mp}{M_{\text{P}}}
\DeclareMathOperator{\Met}{Met}

\begin{document}
\title{Canonical variational completion and 4D Gauss-Bonnet gravity}
\author{Manuel HOHMANN, Christian PFEIFER\inst{1} \and Nicoleta VOICU\inst{2}
}                     
\institute{Laboratory of Theoretical Physics, Institute of Physics, University of Tartu, W. Ostwaldi 1, 50411 Tartu, Estonia \and Faculty of Mathematics and Computer Science, Transilvania University, Iuliu Maniu Str. 50, 500091 Brasov, Romania}
\date{Received: date / Revised version: date}
%
\abstract{
Recently, a proposal to obtain a finite contribution of second derivative order to the gravitational field equations in \(D = 4\) dimensions from a renormalized Gauss-Bonnet term in the action has received a wave of attention. It triggered a discussion whether the employed renormalization procedure yields a well-defined theory. One of the main criticisms is based on the fact that the resulting field equations cannot be obtained as the Euler-Lagrange equations from a diffeomorphism invariant action. In this work, we use techniques from the inverse calculus of variations to point out that the renormalized truncated Gauss-Bonnet equations cannot be obtained from any action at all (either diffeomorphism invariant or not), in any dimension.
Then, we employ canonical variational completion, based on the notion of Vainberg-Tonti Lagrangian - which consists in adding a canonically defined correction term to a given system of equations, so as to make them derivable from an action. To apply this technique to the suggested $4$D renormalized Gauss-Bonnet equations, we extend the variational completion algorithm to some classes of PDE systems for which the usual integral providing the Vainberg-Tonti Lagrangian diverges. We discover that in $D>4$ the suggested field equations can be variationally completed, choosing either the metric or its inverse as field variables; both approaches yield consistently the same Lagrangian, whose variation leads to fourth order field equations. In $D=4$, the Lagrangian of the variationally completed theory diverges in both cases.
} 
\maketitle
%


\section{Introduction}\label{sec:intro}
The study of alternative and extended theories of gravity besides general relativity is motivated by observations in cosmology, such as the accelerating expansion of the universe, and by its tension with quantum theory. The latter has stipulated to consider quantum corrections to the Einstein-Hilbert action in form of higher curvature invariants. One such invariant, which is purely topological in four spacetime dimensions, is the Gauss-Bonnet invariant. Even though it does not contribute to the gravitational field equations in four dimensions, it has been shown that the contributions arising in higher dimensions can be renormalized in such a way as to yield a non-trivial contribution also in the limit of four dimensions~\cite{Tomozawa:2011gp,Cognola:2013fva}. Considering these higher curvature terms as new terms in a classical theory of gravity instead of quantum corrections has led to the proposal of a new ``renormalized 4D Gauss-Bonnet'' theory gravity, which claims to yield a finite second derivative order contribution to the Einstein equations from a Gauss-Bonnet term in the gravitational action~\cite{Glavan:2019inb}. It is thus based on a similar concept as a dimensional regularization of the Einstein-Hilbert action, which is topological in \(D = 2\) dimensions~\cite{Mann:1991qp}. Also a generalization to further Lovelock curvature terms appearing at higher dimensions has been considered~\cite{Casalino:2020kbt}.

However attractive, the proposed model has also received criticism, and the correctness of the procedure of obtaining field equations (or even symmetric solutions) from a renormalization of the Gauss-Bonnet term in the gravitational action so as to obtain a finite, non-vanishing contribution in the limit of \(D \to 4\) dimensions, has been challenged. Various works have shown inconsistencies in the proposed approach. The most obvious contradiction, which also concerns the aforementioned regularized theories in other dimensions, arises from the fact that the existence of a theory with the claimed properties would violate Lovelock's result \cite{Lovelock:1969} that the only generally covariant Lagrangian field theory of the metric tensor alone, giving second-order field equations in four dimensions, is given by general relativity, possibly with a cosmological constant~\cite{Lovelock:1969}. One manifestation of this contradiction is the ``index problem'', which states that certain terms in the field equations vanish due to the number of possible index combinations in a given dimension - which is a discrete number, and therefore does not allow for a continuous limiting procedure. This has been pointed out already for the case of \(D = 2\) regularized Einstein gravity~\cite{Ai:2020peo}, and applies also to the proposed 4D Gauss-Bonnet theory~\cite{Gurses:2020ofy,Mahapatra:2020rds}. Further, it has been shown that the proposed model yields consistent solutions only for highly symmetric spacetimes~\cite{Tian:2020nzb}, and that the obtained regularized field equations cannot be obtained from a regular, diffeomorphism-invariant action~\cite{Arrechea:2020evj}.

In order to circumvent the aforementioned shortcomings and to obtain a consistent 4-dimensional theory preserving certain features of the proposed Gauss-Bonnet theory, various approaches have been studied. One possible approach is to consider the 4D Gauss-Bonnet theory as arising from the Kaluza-Klein reduction of a higher-dimensional theory. This approach leads to the appearance of a scalar Kaluza-Klein mode, which introduces additional contributions to the gravitational field equations also in 4 dimensions, which then reproduce the proposed contribution from the Gauss-Bonnet term~\cite{Kobayashi:2020wqy,Lu:2020iav}. Another approach that yields a supplementary scalar mode is by introducing a counter-term in the gravitational action, which is constructed from a conformally rescaled metric~\cite{Hennigar:2020lsl,Fernandes:2020nbq,Easson:2020mpq}, in analogy to a similar procedure for \(D = 2\) Einstein gravity~\cite{Mann:1992ar}. These extensions are in line with Lovelock's theorem, as they introduce another dynamical field besides the metric, and so it is possible to obtain second-order, local, Lagrangian field equations in 4 dimensions. Hence, they fall into the Horndeski class of gravity theories~\cite{Horndeski:1974wa}. The resulting scalar-tensor field equations, however, are not equivalent to the originally proposed ones~\cite{Lin:2020kqe}. Other possibilities include: explicitly breaking the invariance of the theory under diffeomorphisms, such as by deriving the field equations from a Hamiltonian approach~\cite{Aoki:2020lig}, or employing holography as a means to obtain non-trivial contributions from boundary terms~\cite{Alkac:2020zhg}.

Both the originally proposed 4D Gauss-Bonnet theory and its scalar-tensor regularizations have received remarkable attention; in particular, highly symmetric solutions to the proposed field equations have been studied, such as black holes~\cite{Konoplya:2020bxa,Guo:2020zmf,Fernandes:2020rpa,Konoplya:2020qqh,Wei:2020ght,Kumar:2020owy,Hegde:2020xlv,Ghosh:2020vpc,Zhang:2020qew,Singh:2020xju,Konoplya:2020ibi,Ghosh:2020syx,Konoplya:2020juj,Zhang:2020qam,HosseiniMansoori:2020yfj,Kumar:2020uyz,Roy:2020dyy,Singh:2020nwo,Wei:2020poh,Churilova:2020aca,Kumar:2020xvu,Islam:2020xmy,Mishra:2020gce,Liu:2020vkh,Konoplya:2020cbv,Jin:2020emq,Heydari-Fard:2020sib,Zhang:2020sjh,EslamPanah:2020hoj,NaveenaKumara:2020rmi,Aragon:2020qdc,Yang:2020czk,Cuyubamba:2020moe,Ying:2020bch,Rayimbaev:2020lmz,Liu:2020evp,Zeng:2020dco,Ge:2020tid,Hennigar:2020fkv,Kumar:2020sag,Ghosh:2020cob,Churilova:2020mif,Yang:2020jno,Devi:2020uac,Jusufi:2020qyw,Konoplya:2020der,Liu:2020lwc,Qiao:2020hkx,Dadhich:2020ukj,Shaymatov:2020yte,Hennigar:2020zif,Singh:2020mty,Malafarina:2020pvl}, wormholes~\cite{Jusufi:2020yus,Liu:2020yhu,Chew:2020lkj}, other compact objects~\cite{Doneva:2020ped,Malafarina:2020pvl,Banerjee:2020stc,Banerjee:2020yhu} and cosmology~\cite{Li:2020tlo,Ma:2020ufk,Samart:2020sxj,Narain:2020qhh,Aoki:2020iwm,MohseniSadjadi:2020jmc}. Bounds on the theory have been obtained from its weak field limit~\cite{Clifton:2020xhc} as well as cosmological perturbations~\cite{Casalino:2020pyv,Haghani:2020ynl} and the speed of gravitational waves~\cite{Aoki:2020iwm,Feng:2020duo}. Also the asymptotic structure~\cite{Lu:2020mjp} as well as aspects of quantum gravity~\cite{Shu:2020cjw,Bonifacio:2020vbk} and quantum cosmology~\cite{Narain:2020tsw} have been studied. \\

For field equations which are not variational, one may seek for alternative approaches to finding a set of regular, variational field equations, without the explicit introduction of a scalar degree of freedom. In particular, one may pose the question which set of variational field equations for the metric tensor alone would be as close as possible to the proposed equations. A constructive approach to answer this question is the method of \textit{canonical variational completion}~\cite{Voicu:2015dxa,Krupka-book}. Starting from an arbitrary set of differential equations defined on a specific coordinate chart, it yields a Lagrangian on the respective coordinate chart, whose Euler-Lagrange equations coincide with the original set of differential equations if and only if these are variational. In case they are not, it gives a canonical, in a sense minimal, correction term to be added to the original equations, such that they become variational. A standard example is provided by the Ricci tensor, whose canonical variational completion is the full Einstein tensor. Another example of  successful variational completion is the canonical variational field equation for Finsler gravity~\cite{Hohmann:2019sni}.

In the paper, we show that the truncated Gauss-Bonnet gravity equations cannot be obtained by varying any action (coordinate-invariant or not). Moreover, we study the possibility of extending the proposed field equations for 4-dimensional Gauss-Bonnet gravity using the method of canonical variational completion. To do this consistently, we first extend the construction of the Vainberg-Tonti Lagrangian - which lies at the heart of variational completion algorithm - to cases when the usual construction cannot be applied (as the integral providing it diverges); this extended algorithm is then applied to Gauss-Bonnet gravity.

In the case of the truncated Gauss-Bonnet gravity field equations in $D$ dimensions (which, as stated above, are not variational), we derive a Lagrangian whose Euler-Lagrange equations are the canonical variational completion of the original equations. It turns out that these canonically extended equations cannot be of second order in any dimension, and do not even exist in $D=4$.

For this purpose, we split the contribution to the gravitational field equations arising from the Gauss-Bonnet term, into two parts: one part which vanishes identically in \(D = 4\) dimensions for combinatorial reasons, and does not allow for a limit \(D \to 4\), and a part which is proportional to \(D - 4\), and can hence be renormalized to yield a finite contribution also in \(D = 4\) dimensions. We then apply the method of variational completion to each of these terms separately, and demonstrate that due to their degree of homogeneity in the dynamical variables of the theory, the obtained canonical correction still diverges in dimension $D=4$.

The outline of this article is as follows. In Section~\ref{sec:4dgaussbonnet}, we briefly review the proposed field equations of 4D Gauss-Bonnet gravity. A brief review and an extension of the method of canonical variational completion is provided in Section~\ref{sec:varcompl}. We then apply the method to 4D Gauss-Bonnet gravity in Section~\ref{sec:varco4dgb} and end with a conclusion in Section~\ref{sec:conclusion}. Technical details on the derivative order of the variational completion of the truncated field equations are presented in Appendix  \ref{App:A}. Appendix \ref{App:B} contains technical details on the mathematical nature of Euler-Lagrange expressions in the variational completion algorithm. In Appendix \ref{App:C}, we provide the mathematical proofs for the extension of the variational completion algorithm to a whole class of PDE systems for which the usual integral providing the Vainberg-Tonti Lagrangian diverges.

\section{4D Gauss-Bonnet gravity}\label{sec:4dgaussbonnet}
The proposed 4-dimensional extension of Gauss-Bonnet gravity is based on the $D$-dimensional action~\cite{Glavan:2019inb}
\begin{equation}\label{eq:gbaction}
S = \int\dd^Dx\sqrt{-g}\left[\frac{\Mp^2}{2}R - \Lambda_0 + \frac{\alpha}{D - 4}\mathcal{G}\right] + S_m\,,
\end{equation}
where the Gauss-Bonnet scalar is given by
\begin{equation}
\mathcal{G} = 6R^{\mu\nu}{}_{[\mu\nu}R^{\rho\sigma}{}_{\rho\sigma]} = R^2 - 4R_{\mu\nu}R^{\mu\nu} + R_{\mu\nu\rho\sigma}R^{\mu\nu\rho\sigma}\,.
\end{equation}
By variation with respect to the metric \(g_{\mu\nu}\) one obtains the field equations
\begin{equation}\label{eq:gbfield}
E_{\mu\nu} = \Mp^2G_{\mu\nu} + \Lambda_0g_{\mu\nu} - \frac{2\alpha}{D - 4}\mathcal{G}_{\mu\nu} = T_{\mu\nu}\,,
\end{equation}
where
\begin{equation}
G_{\mu\nu} = R_{\mu\nu} - \frac{1}{2}Rg_{\mu\nu}
\end{equation}
is the Einstein tensor, and the term originating from the Gauss-Bonnet scalar is given by
\begin{equation}
\mathcal{G}_{\mu\nu} = 15g_{\mu[\nu}R^{\rho\sigma}{}_{\rho\sigma}R^{\omega\tau}{}_{\omega\tau]} = \frac{1}{2}\mathcal{G}g_{\mu\nu} - 2R_{\mu\lambda\rho\sigma}R_{\nu}{}^{\lambda\rho\sigma} + 4R_{\mu\rho\nu\sigma}R^{\rho\sigma} + 4R_{\mu\rho}R_{\nu}{}^{\rho} - 2RR_{\mu\nu}\,.
\end{equation}
It has been argued in~\cite{Glavan:2019inb} that since \(\mathcal{G}_{\mu\nu} = 0\) in \(D = 4\) dimensions, this theory has a well-defined limit for \(D \to 4\). However, this is a fallacy, since it can be shown that the latter term is given by~\cite{Gurses:2020ofy,Arrechea:2020evj}
\begin{equation}
-\mathcal{G}_{\mu\nu} = (D - 4)A_{\mu\nu} + W_{\mu\nu}\,,
\end{equation}
where we introduced the tensors
\begin{equation}
A_{\mu\nu} = \frac{D - 3}{(D - 2)^2}\left[\frac{2D}{D - 1}RR_{\mu\nu} - 4\frac{D - 2}{D - 3}R^{\rho\lambda}C_{\mu\rho\nu\lambda} - 4R_{\mu}{}^{\rho}R_{\nu\rho} + 2R_{\rho\lambda}R^{\rho\lambda}g_{\mu\nu} - \frac{1}{2}\frac{D + 2}{D - 1}R^2g_{\mu\nu}\right]
\end{equation}
and
\begin{equation}
W_{\mu\nu} = 2C_{\mu}{}^{\rho\lambda\sigma}C_{\nu\rho\lambda\sigma} - \frac{1}{2}C_{\tau\rho\lambda\sigma}C^{\tau\rho\lambda\sigma}g_{\mu\nu}\,,
\end{equation}
using the Weyl tensor
\begin{equation}
C_{\mu\nu\rho\sigma} = R_{\mu\nu\rho\sigma} + \frac{1}{D - 2}(R_{\mu\sigma}g_{\nu\rho} - R_{\mu\rho}g_{\nu\sigma} + R_{\nu\rho}g_{\mu\sigma} - R_{\nu\sigma}g_{\mu\rho}) + \frac{1}{(D - 1)(D - 2)}R(g_{\mu\rho}g_{\nu\sigma} - g_{\mu\sigma}g_{\nu\rho})\,,
\end{equation}
and that one cannot extract a factor \(D - 4\) from the latter term \(W_{\mu\nu}\), which vanishes in \(D = 4\) dimensions for combinatorial reasons. Hence, the field equations~\eqref{eq:gbfield}, which now take the form
\begin{equation}\label{eq:gbfield2}
E_{\mu\nu} = \Mp^2G_{\mu\nu} + \Lambda_0g_{\mu\nu} + 2\alpha\left(A_{\mu\nu} + \frac{W_{\mu\nu}}{D - 4}\right) = T_{\mu\nu}\,,
\end{equation}
do not have a smooth limit for \(D \to 4\), due to the appearance of the last term. It has thus been argued that this term should be omitted in \(D = 4\) dimensions, and only the truncated part \(A_{\mu\nu}\) be considered as the \(D \to 4\) limit of the field equations. However, as shown in~\cite{Arrechea:2020evj}, this term cannot originate from the variation of a diffeomorphism-invariant action. In the following we will show that even dropping the diffeomorphism invariance request, this term cannot originate from any action at all and also the variational completion of these truncated field equations degenerates in $D=4$.

\section{Variational completion}\label{sec:varcompl}
We first briefly review the method of canonical variational completion~\cite{Voicu:2015dxa,Krupka-book} and introduce an extension thereof, which we will use in the next section. Given an arbitrary PDE system, the inverse variational problem consists in finding out whether there exists a Lagrangian function $\mathcal{L}$ having the PDE system as its Euler-Lagrange equations.

\subsection{Classical variational completion}
In the following, we will present for simplicity, the case of second order PDE systems (yet, the theory works for equations of arbitrary order):
\begin{equation}
\mathcal{E}_{A}(x^{\mu },y^{B},y_{~\mu }^{B},y_{~\mu \nu }^{B})=0,
\label{PDE}
\end{equation}
where $x^\mu\ (\sigma ,\mu, \nu =0,...,n-1)$ are coordinates on a smooth manifold $M$, $y^A\ (A,B=1,...,m)$ are functions of $x^\mu$ and subscripts in $y^A{}_{\mu} = \partial{}_{\mu} y^A$  $y^A{}_{\mu\nu} = \partial_{\mu} \partial_{\nu}  y^A$ stand for partial differentiation.

The dependent variables $y^A=y^A(x^{\mu})$ will be regarded as components of \textit{sections} $\gamma:U \rightarrow Y, (x^{\mu}) \mapsto y^{A}(x^{\mu})$ (where $U \subset M$ is open) into fiber bundles $(Y\overset{\pi}{\rightarrow}M,F)$ over $M$; in the context of physics, sections into fiber bundles are interpreted as physical fields.\\
Also, we will consider $\mathcal{E}_{A}$ as functions defined on some fibered coordinate chart $(V^{2},\psi ^{2})$ on the second order jet bundle $J^{2}Y$.
Concretely, going to $J^{2}Y$ allows us to treat $x^{\mu},y^{B},y_{~\mu }^{B},y_{~\mu \nu }^{B}$ with $\mu \leq \nu,$ as (independent) coordinate functions associated to a chart $(V^{2},\psi ^{2})$ - and, e.g., to rigorously define partial derivatives with respect to them.
Intuitively, $x^{\mu},y^{B},y_{~\mu }^{B},y_{~\mu \nu }^{B}$ play the role of independent "slots" into which, when we insert sections $y^A=y^A(x^{\mu})$, they return the dependent variables of our PDE system and their partial derivatives.

Also,  Lagrangians will be regarded as differential forms on $V^{2}$:
\begin{equation*}
\lambda =\mathcal{L}\dd^{n}x,
\end{equation*}%
where $\mathcal{L} = \mathcal{L}(x^{i},y^{B},y_{~\mu }^{B},y_{~\mu \nu }^{B})$. \\

The system (\ref{PDE}) is called:
\begin{itemize}
	\item \textit{locally variational} if, corresponding to any given fibered chart $%
	(V^{2},\psi ^{2})$ on $J^2Y$, there exists a Lagrangian $\lambda _{V}$ on $%
	V^{2}$ having (\ref{PDE}) as its Euler-Lagrange equations;
	\item \textit{globally variational} if (\ref{PDE}) admits a Lagrangian $\lambda
	=\mathcal{L}\dd^{n}x$ defined on the entire $J^{2}Y,$ i.e., the various
	Lagrangians $\lambda _{V}$ defined on each fibered chart of $J^{2}Y,$ can be smoothly glued together into a single
	Lagrangian $\lambda$.
\end{itemize}

In the following, by "variational", unless elsewhere specified, we will mean locally variational. It is important to note that, here, the term "local" means "defined over a specific coordinate chart", i.e., it
has a different meaning than the one it commonly has in physics. It does neither imply that the Lagrangian is coordinate invariant, nor that it might not contain, e.g., integrals.

There are basically two ways of checking whether a given PDE system is locally variational: checking the so-called Helmholtz conditions, \cite{Krupka-book}, or explicitly finding a Lagrangian. These two methods are tightly related, as follows.


Given the PDE system \eqref{PDE} and assuming that $(V^{2},\psi ^{2})$ is \textit{vertically star-shaped}, i.e.\ for every $(x^{\mu },y^{B},y_{~\mu }^{B},y_{~\mu \nu }^{B})\in \psi^{2}(V^{2})$ also the whole segment $(x^{\mu },t y^{B},t y_{~\mu }^{B},t y_{~\mu \nu }^{B}), \textrm{ for } t\in[0,1]$, remains in $\psi^{2}(V^{2})$, we introduce the so-called \textit{Vainberg-Tonti Lagrangian} function $\mathcal{L}$ on the given coordinate neighborhood $V^{2}$, by:
\begin{equation}
\mathcal{L}_{\mathcal{E} }(x^{\mu},y^{B},y^{B}{}_{\mu },y^{B}{}_{\mu \nu })=y^{A}\int_{0}^{1}\mathcal{E}_{A}(x^{\mu
},ty^{B},ty^{B}{}_{\mu },ty^{B}{}_{\mu \nu })\dd t.  \label{eq:vtlag}
\end{equation}%
Under the above assumption on $V^{2}$, the Vainberg-Tonti Lagrangian is well-defined and gives rise to the Euler-Lagrange expressions%
\begin{equation}
\tilde{\mathcal{E}}_{A}:=\frac{\partial \mathcal{L}_{\mathcal{E} }}{\partial y^{A}}-\mathrm{%
	d}_{\mu }\frac{\partial \mathcal{L}_{\mathcal{E} }}{\partial y^{A}{}_{\mu }}+
\mathrm{d}_{\mu }\mathrm{d}_{\nu }\frac{\partial \mathcal{L}_{\mathcal{E} }}{\partial
	y^{A}{}_{\mu \nu }}\,,  \label{eq:eulag}
\end{equation}
where $\dd_\mu = \frac{\dd}{\dd x^\mu}$.

An important result from variational calculus states that a system of
partial differential equations for which the integral \eqref{eq:vtlag} makes
sense is locally variational if and only if the obtained Euler-Lagrange
equations $\tilde{\mathcal{E}}_{A}=0$ coincide with the original equations~%
\eqref{PDE}. Actually, it can be shown that the correction terms
\begin{equation}
H_{A}=\tilde{\mathcal{E}}_{A}-\mathcal{E}_{A}  \label{eq:helmholtz}
\end{equation}%
are linear combinations of the coefficients of the so-called \textit{Helmholtz form} - see (\ref{HA}) and (\ref{H1})-(\ref{H3}) in the Appendices below for their precise expressions. If the given equations are locally variational, then the associated Helmholtz form identically vanishes. Conversely, on vertically star-shaped domains $V^{2}$, the vanishing of the Helmholtz coefficients implies the local variationality of $\mathcal{E}_{A}=0$; more precisely,
the Vainberg-Tonti Lagrangian is a Lagrangian for \eqref{PDE} and any other Lagrangian for  \eqref{PDE} will differ from the Vainberg-Tonti Lagrangian by a	divergence expression - which will thus bring no contribution to the Euler-Lagrange equations, see, e.g., \cite{Krupka-book}.

In other words, the mapping attaching to a class of equivalent Lagrangians, their common Euler-Lagrange expressions (more technically, their common $\textit{Euler-Lagrange source form}$), and the mapping attaching to a given variational source form (i.e., to a set of Euler-Lagrange expressions), its Vainberg-Tonti Lagrangian, are inverse to each other.

The system~\eqref{eq:eulag} of partial differential equations, which is locally variational by construction, is called the \textit{canonical variational completion} of the original system~\eqref{PDE}. Thus, the canonical variational completion of \eqref{PDE} is obtained by adding the corresponding Helmholtz expressions; if \eqref{PDE} is not variational, then, these provide nontrivial correction terms to the original system. A standard example in this sense is the following. \\

\textbf{Example: the Einstein tensor as the canonical variational completion of the Ricci tensor \cite{Voicu:2015dxa}.} \\
Historically, it is known that the first variant of gravitational field equations proposed by Einstein was
\begin{equation}\label{eq:oldefe}
\Mp^2 R_{\mu \nu} = T_{\mu \nu}.
\end{equation}
Later on, he noticed that this system was inconsistent, since the right hand side is covariantly divergence-free, while the Ricci tensor is not; based on a heuristic argument involving the contracted Bianchi identity, he then added the correction term $-\frac{1}{2} R g_{\mu \nu}$ to the left hand side, thus obtaining the nowadays known form of the fundamental equations of general relativity.

Yet, there is another reasoning, based on the calculus of variations, which leads to the same conclusion. The system \eqref{eq:oldefe} is not variational; again, the one at fault is the left hand side, which cannot be obtained as the Euler-Lagrange equation attached to any Lagrangian defined on the given coordinate neighborhood. A straightforward application of the outlined canonical variational completion algorithm  to the tensor density $\sqrt{-g}R^{\mu \nu}$ (considering the components $g_{\mu\nu}$ of the metric as fundamental variables) gives the Vainberg-Tonti Lagrangian function $\mathcal{L}_{\mathcal{E} } = R \sqrt{-g}$. It leads to the variationally completed field equations - which are the Einstein field equations as nowadays known.

\subsection{Modified variational completion}\label{ssec:extVC}

The assumption on the vertically star-shapedness of the coordinate chart $(V^2,\psi^2)$ puts a limitation on the applicability of the the Vainberg-Tonti Lagrangian \eqref{eq:vtlag}.
As an immediate example, in metric theories where the dynamical variables are the metric components $g_{\mu\nu}$ or the inverse metric components $g^{\mu \nu}$, the coordinate neighborhood $V^{2}$ is not vertically star-shaped; to be more specific, the lower integration endpoint $t=0$ in the Vainberg-Tonti Lagrangian corresponds to $g_{\mu\nu} = 0$ (or, accordingly, to $g^{\mu \nu} =0$) - but none of these corresponds to a well-defined metric tensor. \\
The first idea which comes to mind, in such cases, is to consider the lower integration endpoint $t=0$ as a \textit{limit}. And, while in some cases, such a limit exists and is finite (as, for instance, in the case of the Ricci tensor, presented above), so we can still define the Vainberg-Tonti Lagrangian by the same formula - in some other cases, the resulting improper intergal diverges. This is, e.g., the case of
PDE systems $\mathcal{E}_{A}(x^{\mu},ty^{B},ty^{B}{}_{\mu },ty^{B}{}_{\mu \nu })=0$ that are homogeneous of negative degree smaller or equal to $-1$ in the fiber variables $y^{B}.$ In these cases, not only the lower integration endpoint $t=0$ does not correspond to points in the domain of definition of the functions $\mathcal{E}_{A}$ - but due to it, the integral providing the Vainberg-Tonti Lagrangian becomes infinite. \\

A way of extending the applicability of the variational completion algorithm is to replace the integration endpoint $t=0$ with a more general one. More precisely, assume that the functions $\mathcal{E}_{A}$ are defined over some arbitrary fibered chart domain $V^{2} \subset J^{2}Y$. We will prove the following

\begin{theorem}\label{VT-extended}
	Let $\mathcal{E}_{A} = 0$ be an arbitrary second order PDE system and $a \in \mathbb{R} \cup \{\pm \infty\}$ such that
	\[
	\underset{t\rightarrow a}{\lim }\left( t\mathcal{E} _{B}(x^{\mu
	},ty^{A},ty_{~\mu }^{A},ty_{~\mu \nu }^{A})\right) =0\,
	\]%
	for all $(x^{\mu },y^{B},y_{~\mu }^{B},y_{~\mu \nu }^{B})$ in the domain of definition of $\mathcal{E}_{A}$. Define, at these points, the extended Vainberg-Tonti Lagrangian $\lambda = \mathcal{L}_{\mathcal{E}} \dd^{n}x$, by the rule:
	\begin{equation}\label{eq:VText}
	\mathcal{L}_{\mathcal{E} }(x^{\mu },y^{B},y_{~\mu }^{B},y_{~\mu \nu }^{B}):=y^{A}\overset{1}{\underset{a}{\int }}\mathcal{E}
	_{A}(x^{\mu },ty^{B},ty_{~\mu }^{B},ty_{~\mu \nu }^{B})\dd t\,.
	\end{equation}%
	If the above integrals exist and are finite, then:\\
	1. If the equations $\mathcal{E}_{A} = 0$ are variational, then $\lambda$ is a (locally defined) Lagrangian for these, i.e., the Euler-Lagrange expressions of \eqref{eq:VText} are precisely $\mathcal{E}_{A}$:
	\begin{equation}
	\tilde{\mathcal{E}}_{A} =\mathcal{E} _{A}.
	\end{equation}

	2. If $\mathcal{E}_{A} = 0$ are not variational, then the Euler-Lagrange expressions of \eqref{eq:VText} are their canonical variational completion; the correction terms are expressed in terms of the coefficients $H_{AB}, H^{\mu}_{AB}, H^{\mu \nu}_{AB}$ of the Helmholtz form:
	\begin{equation}
	\tilde{\mathcal{E}}_{A} = \mathcal{E} _{A} + H_A,
	\end{equation}
	where
	\begin{equation}\label{H_A}
	H_A =  -\overset{b}{\underset{a}{\int }} t[y^{B}(H_{AB}\circ \chi _{t})+y_{~\mu
	}^{B}(H_{AB}^{\mu }\circ \chi _{t})+y_{~\mu \nu }^{B}(H_{AB}^{\mu \nu }\circ
	\chi _{t})] \dd t.
	\end{equation}

\end{theorem}
The proof of this theorem can be found in Appendix \ref{App:C} and the precise expressions of the Helmholtz form coefficients  $H_{AB}, H^{\mu}_{AB}, H^{\mu \nu}_{AB}$ are presented in Appendix \ref{App:B}.

\bigskip

In the particular case $a=0$, the above formula gives the usual Vainberg-Tonti Lagrangian. \\

This result will be used when we apply the variational completion to 4D Gauss-Bonnet gravity - and it ensures that our conclusion does not depend on the choice of $g_{\mu\nu}$ or $g^{\mu\nu}$ as dynamical variables.

\section{Variational completion of 4D Gauss-Bonnet gravity}\label{sec:varco4dgb}

Let us start by the following: \\
\textbf{Remark.} \textit{The truncated Gauss-Bonnet equations cannot be obtained as the Euler-Lagrange equations of any Lagrangian (coordinate invariant or not)}. \\

\bigskip

To justify the above statement, we use the following result, \cite[p. 147]{Krupka-book}: If a second order PDE system is locally variational, then it must be linear in the second order derivatives of the dependent variable. Or, the truncated Gauss-Bonnet terms $A_{\mu \nu}$ are not linear in the second order derivatives of $g_{\mu \nu}$, see Appendix \ref{App:A}, in any dimension. Hence, independently of the value of $D$, they cannot represent the Euler-Lagrange expressions of any Lagrangian - either coordinate invariant or not. \\

In the following, we aim to determine a correction term to be added to $A_{\mu \nu}$, so as to make them variational.

More precisely, we will apply the above described canonical variational completion algorithm to the truncated Gauss-Bonnet gravity equations
\begin{align}\label{eq:GBtrunc}
\mathring{E}_{\mu\nu} = \Mp^2G_{\mu\nu} + \Lambda_0g_{\mu\nu} + 2\alpha A_{\mu\nu} = T_{\mu\nu}\,,
\end{align}
where the over-set circle denotes truncation.
\\

As a preliminary step, let us first check the Vainberg-Tonti Lagrangian for the full, non-truncated the field equations ~\eqref{eq:gbfield2}; since they are, by construction, variational, we should obtain - at least, up to a divergence expression - the original Lagrangian \eqref{eq:gbaction}. In order to use the components of the metric $g_{\mu\nu}$ as our dynamical variables $y^A$, we need to raise the indices and to restore the density factor in the field equations~\eqref{eq:gbfield}, such that they read
\begin{equation}\label{eq:densfeq}
\mathcal{E}^{\mu\nu} = -\frac{1}{2}\sqrt{-g}E^{\mu\nu}\,,
\end{equation}
where the factor \(-\frac{1}{2}\) arises from the definition of the energy-momentum tensor in the field equations~\eqref{eq:gbfield}. Hence, these are the correct Euler-Lagrange expressions obtained from variation of the action \eqref{eq:gbaction}. Without this density factor, the equations cannot be variational, which can be proven using the Helmholtz conditions - see Appendix~\ref{App:B} for the mathematical details. Thus, omitting the density would break the relation between the Vainberg-Tonti Lagrangian of an already variational system, and the variation, which recovers these equations as its Euler-Lagrange equations. \\

Note that under a rescaling \(g_{\mu\nu} \to tg_{\mu\nu}\), the terms in the field equations \eqref{eq:densfeq} transform as
\begin{equation}
g_{\mu\nu} \to tg_{\mu\nu}\,, \quad
G_{\mu\nu} \to G_{\mu\nu}\,, \quad
A_{\mu\nu} \to t^{-1}A_{\mu\nu}\,, \quad
W_{\mu\nu} \to t^{-1}W_{\mu\nu}\,,
\end{equation}
so that after raising indices and restoring the density factor we have
\begin{equation}
g^{\mu\nu} \to t^{-1}g^{\mu\nu}\,, \quad
\sqrt{-g} \to t^{\frac{D}{2}}\sqrt{-g}\,, \quad
G^{\mu\nu} \to t^{-2}G^{\mu\nu}\,, \quad
A^{\mu\nu} \to t^{-3}A^{\mu\nu}\,, \quad
W^{\mu\nu} \to t^{-3}W^{\mu\nu}\,.
\end{equation}

Since the degree of homogeneity of the functions $\mathcal{E}^{\mu\nu}$ with respect to $g_{\mu\nu}$ is larger than $-1$ in $D>4$, we can apply the classical Vainberg-Tonti Lagrangian \eqref{eq:vtlag} to obtain
\begin{equation}\label{eq:vainbergtonti}
\begin{split}
\mathcal{L} &= -\frac{1}{2}g_{\mu\nu}\int_0^1t^{D/2}\sqrt{-g}\left[t^{-2}\Mp^2G^{\mu\nu} + t^{-1}\Lambda_0g^{\mu\nu} + 2t^{-3}\alpha\left(A^{\mu\nu} + \frac{W^{\mu\nu}}{D - 4}\right)\right]\dd t\\
&= -\frac{1}{2}\sqrt{-g}g_{\mu\nu}\left[\frac{2\Mp^2}{D - 2}G^{\mu\nu} + \frac{2\Lambda_0}{D}g^{\mu\nu} + \frac{4\alpha}{D - 4}\left(A^{\mu\nu} + \frac{W^{\mu\nu}}{D - 4}\right)\right]\\
&= \sqrt{-g}\left[\frac{\Mp^2}{2}R - \Lambda_0 - \frac{2\alpha}{D - 4}\left(A^{\mu}{}_{\mu} + \frac{W^{\mu}{}_{\mu}}{D - 4}\right)\right]\,,
\end{split}
\end{equation}
where we used \(g_{\mu\nu}g^{\mu\nu} = D\) as well as
\begin{equation}
g_{\mu\nu}G^{\mu\nu} = g_{\mu\nu}\left(R^{\mu\nu} - \frac{1}{2}Rg^{\mu\nu}\right) = R - \frac{D}{2}R = \left(1 - \frac{D}{2}\right)R\,,
\end{equation}
and the appearing traces are given by
\begin{align}
A^{\mu}{}_{\mu} &= \frac{D - 3}{D - 2}\left(2R_{\mu\nu}R^{\mu\nu} - \frac{D}{2(D - 1)}R^2\right)\,,\\
W^{\mu}{}_{\mu} &= (D - 4)\left(\frac{2}{D - 2}R_{\mu\nu}R^{\mu\nu} - \frac{1}{(D - 1)(D - 2)}R^2 - \frac{1}{2}R_{\mu\nu\rho\sigma}R^{\mu\nu\rho\sigma}\right)\,.
\end{align}
We see that the first two terms in the last line of \eqref{eq:vainbergtonti} give us the Einstein-Hilbert Lagrangian density plus the densitized cosmological constant, as one would have expected. For the last terms, however, we observe the following:

\begin{enumerate}
	\item
	First, in \(D = 4\) dimensions, the procedure giving us the Vainberg-Tonti Lagrangian density is not well-defined for any of the last two terms (taken separately), since in that case one would obtain a term proportional to \(t^{-1}\) in the integral~\eqref{eq:vainbergtonti} - which would thus diverge. Hence, we can perform the integration only in \(D > 4\) dimensions, in which case one obtains a factor \((D - 4)^{-1}\) in the result.

	\item
	The combination of the last two terms satisfies
	\begin{equation} \label{combi}
	A^{\mu}{}_{\mu} + \frac{W^{\mu}{}_{\mu}}{D - 4} = -\frac{1}{2}\mathcal{G}\,,
	\end{equation}
	and so the full Lagrangian~\eqref{eq:vainbergtonti} indeed recovers the gravitational part of the action~\eqref{eq:gbaction}, hence confirming the validity of the variational completion procedure for the full field equations~\eqref{eq:gbfield2}.

	\item The Vainberg-Tonti Lagrangian of the truncated field equations~\eqref{eq:GBtrunc}, with source form $\mathring{\mathcal{E}}^{\mu\nu} = -\frac{1}{2}\sqrt{-g}\mathring{E}^{\mu\nu}$, in which the term \(W_{\mu\nu}\) has been omitted, yields the truncated Lagrangian
	\begin{align}
	\mathring{\mathcal{L}} = \sqrt{-g}\left[\frac{\Mp^2}{2}R - \Lambda_0 - \frac{2\alpha}{D - 4} A^{\mu}{}_{\mu} \right]\,,
	\end{align}
	which is well defined in any dimension $D$ except $D=4$. Its variation does not give back the original truncated field equations~\eqref{eq:GBtrunc}, but their canonical variationally completed field equations, which have been obtained using the Mathematica package xAct \cite{xact} and xTras \cite{xtras},
	\begin{multline}\label{eq:varcompegbtrunc}
	\tilde{\mathring{E}}_{\mu\nu} = \Mp^2G_{\mu\nu} + \Lambda_0g_{\mu\nu} + \frac{4\alpha(D - 3)}{(D - 1)(D - 2)(D - 4)}\bigg[g_{\mu\nu}\left(\Box R - \frac{D}{4}R^2 + (D - 1)R^{\rho\sigma}R_{\rho\sigma}\right)\\
	- 2(D - 1)\Box R_{\mu\nu} + (D - 2)\nabla_{\mu}\nabla_{\nu}R + DRR_{\mu\nu} - 4(D - 1)R^{\rho\sigma}R_{\mu\rho\nu\sigma}\bigg] = T_{\mu\nu}\,.
	\end{multline}
	They clearly contain fourth order derivatives acting on the metric, usually canceled from the contributions coming from the $W^\mu{}_\mu$ term. Also, it becomes explicitly visible that these field equations diverge in $D=4$.
	\item For future investigations of this theory in view of cosmology, AdS/CFT-correspondence and possible emergence from a low energy limit of string theory, we point out that for Einstein spacetimes, in which $R_{\mu\nu} = k g_{\mu\nu}$, where $k$ is a real number, the field equation \eqref{eq:varcompegbtrunc} becomes
	\begin{align}\label{eq:FEQEinsteinSpace}
	\left[\Mp^2 \left(1 - \tfrac{D}{2}\right)k + \Lambda_0 - \alpha k^2 \left(D-4+\tfrac{2}{D-1}\right)\right] g_{\mu\nu} = T_{\mu\nu}\,.
	\end{align}
	Thus, de Sitter and anti-de Sitter spacetime are vacuum solutions of the field equations, with a curvature constant $k$ determined by by $\Lambda_0,D$ and $\alpha$. A thorough and detailed analysis of the properties of this theory is beyond the scope of this article and will be investigated in future research.
\end{enumerate}

\textbf{Remark.} As we have seen, the classical Vainberg-Tonti Lagrangian~\eqref{eq:vainbergtonti} diverges for $D=4$. To ensure that this is not an artifact of our choice of variable, we now consider the components of the inverse metric \(g^{\mu\nu}\) as fundamental fields and employ the extended Vainberg-Tonti Lagrangian \eqref{eq:VText}.

The field equations to be used, together with a proper density factor take the form
\begin{equation}\label{eq:denfeq2}
\mathcal{E}_{\mu\nu} = -\frac{1}{2}\sqrt{-g}E_{\mu\nu}\,,
\end{equation}
where \(E_{\mu\nu}\) is given by~\eqref{eq:gbfield2}. Consider the rescaling \(g^{\mu\nu} \mapsto tg^{\mu\nu}\), under which the terms in the terms in the densitized field equations~\eqref{eq:denfeq2} behave as
\begin{equation}
g_{\mu\nu} \to t^{-1}g_{\mu\nu}\,, \quad
\sqrt{-g} \to t^{-\frac{D}{2}}\sqrt{-g}\,,\quad
G_{\mu\nu} \to G_{\mu\nu}\,, \quad
A_{\mu\nu} \to tA_{\mu\nu}\,, \quad
W_{\mu\nu} \to tW_{\mu\nu}\,.
\end{equation}
For the modified Vainberg-Tonti Lagrangian density one then finds the expression
\begin{equation}
\mathcal{L} = -\frac{1}{2}g^{\mu\nu}\int_a^1t^{-D/2}\sqrt{-g}\left[\Mp^2G_{\mu\nu} + t^{-1}\Lambda_0g_{\mu\nu} + 2t\alpha\left(A_{\mu\nu} + \frac{W_{\mu\nu}}{D - 4}\right)\right]\dd t\,,
\end{equation}
where we have used $a$ instead of $0$ as the lower integration endpoint; due to the appearance of powers \(t^k\) with \(k \leq -1\) (see the second term), \ choosing $a=0$, the integral would diverge for any $D>0$. But, we note that $a=\infty$ satisfies
\[
\underset{t\rightarrow a}{\lim }\left( t\mathcal{E} _{B}(x^{\mu
},ty^{A},ty_{~\mu }^{A},ty_{~\mu \nu }^{A})\right) =0\,,
\]
hence, according to Theorem \ref{VT-extended}, it can be used as an integration endpoint. With this choice, we get again \eqref{eq:vainbergtonti} as the Vainberg-Tonti Lagrangian of the (non-truncated) Gauss-Bonnet field equations. The Vainberg-Tonti Lagrangian of the truncated field equations is obtained by omitting the $W_{\mu\nu}$ term - and, again, it diverges for $D=4$. \\
Hence, no matter whether we use $g_{\mu \nu}$ or $g^{\mu \nu}$ as our dynamical variables, the Vainberg-Tonti Lagrangian for the truncated Gauss-Bonnet field equations diverges in $D=4$.





\section{Summary and outlook}\label{sec:conclusion}
The variational completion algorithm is a powerful tool to answer the question if a certain set of field equations are the Euler-Lagrange equations of an action principle and, in the negative case, it determines a correction term to the original equations, which makes them variational.

The classical variational completion relies on the existence of the Vainberg-Tonti Lagrangian \eqref{eq:vtlag} - which depends on the choice of variables in the PDE system under consideration. For example, in theories of gravity, where the fundamental variables are the components of the spacetime metric or (which is physically equivalent) the components of the inverse spacetime metric, it turned out that, for one choice of the variable, the classical Vainberg-Tonti Lagrangian may exist, while for the other choice, it may very well not exist. Technically, this problem emerges since in metric theories of gravity the values of the field variables (either $g_{\mu\nu}$ or $g^{\mu\nu}$) are not in a star-shaped domain with center zero. To extend the applicability of the Vainberg-Tonti Lagrangian and to ensure that our results on the variationality and the variational completion of the ``renormalized 4D Gauss-Bonnet'' theory of gravity, which was suggested in~\cite{Glavan:2019inb}, are independent of the choice of the metric or its inverse as the dependent variable, we presented an extended formulation of the Vainberg-Tonti Lagrangian in Theorem \ref{VT-extended}. The key insight, to include PDE systems that have negative homogeneity  of degree smaller $-1$ with respect to the fundamental field variable, was to properly adapt the integration endpoints in the Vainberg-Tonti construction.


Applying the extended variational completion to the ``renormalized 4D Gauss-Bonnet theory'', we confirmed from the standpoint of the inverse problem of the calculus of variations, independently of the choice of the metric or inverse metric components as field variables, that the field equations \eqref{eq:GBtrunc} cannot be variational in any dimension. Moreover, we have proven that the variational completed theory diverges in $D = 4$ dimensions, and leads to higher order field equations, which we derived explicitly, in \(D > 4\), see \eqref{eq:varcompegbtrunc}. The reason for this is that the separate terms \(A_{\mu\nu}\) and \(W_{\mu\nu}\) do not just reproduce their individual contribution to the field equations, but also additional terms of fourth derivative order, which would otherwise cancel when they are summed to form the Gauss-Bonnet term.

Having found the variational completion~\eqref{eq:varcompegbtrunc} of the truncated Gauss-Bonnet field equations in $D>4$, the next steps are to study the properties of this theory and its potential for a geometric explanation of dark energy.
In contrast to the well studied Gauss-Bonnet dark energy and inflation scenarios, in which usually scalar fields are added as dynamical couplings to the Gauss-Bonnet term, our newly suggested theory does not contain additional fields to the metric; yet, it introduces instead, higher order derivatives of the metric. Hence, first of all the question of the existence of ghosts has to be investigated, as for example in \cite{Nojiri:2018ouv}. Afterwards it is possible to search for homogeneous and isotropic cosmological solutions of the theory in dimension $D\geq 4$ and study their predictions for the dark energy phenomenology and, in particular for inflation, following for example \cite{Nojiri:2005vv,DeLaurentis:2015fea,Capozziello:2014ioa,Bajardi:2020osh}. One possible outcome is that the solutions obtained will have a well defined $D\to4$ limit and can be interpreted as the geometry of a $4$-dimensional brane embedded in a higher dimensional string space. Another direction of research is to study the AdS/CFT correspondence of the $D$-dimensional AdS solutions, whose curvature parameter is determined by the dimension $D$, the Planck mass $\Mp$, the cosmological constant $\Lambda$ and the Gaus-Bonnet coupling parameter $\alpha$, as we have demonstrated in \eqref{eq:FEQEinsteinSpace}. The corresponding CFT in $D-1$ dimension may reveal the connection of variational completion of the truncated Gauss-Bonnet equations to string theory.

Beyond studying the physical implications of the theory, a mathematical question on the framework of variational completion emerges from our work. We generalized the procedure so as to extend its applicability, for instance to PDE's which are homogeneous of degree smaller than $-1$ in the dependent variable; the question arises if we can further generalize the procedure to PDE systems defined on completely arbitrary, non-star-shaped domains.

\textbf{Acknowledgments.}
	MH and CP were supported by the Estonian Ministry for Education and Science through the Personal Research Funding Grants PRG356 and PSG489, as well as the European Regional Development Fund through the Center of Excellence TK133 ``The Dark Side of the Universe''. The authors would like to acknowledge networking support by the COST Actions CANTATA (CA15117) and QGMM (CA18108), supported by COST (European Cooperation in Science and Technology). This version of the article is the final manuscript draft, it is published in the  \href{https://doi.org/10.1140/epjp/s13360-021-01153-0}{European Physics Journal P 136 (2021)}.

\appendix

\section{Properties of the field equations from the $A^\mu{}_\mu$ term}\label{App:A}
We claimed in item 3 of Section~\ref{sec:varco4dgb} that the variationally completed field equations of the truncated Einstein-Gauss Bonnet field equations \eqref{eq:GBtrunc} contain higher than second order derivatives in any dimension. This is true due to the following line of argument.

The trace of $A_{\mu\nu}$ contains non-trivial terms which are quadratic in the second derivatives of the metric and do not factor in a way that Bianchi identities cancel these terms. Hence also $A_{\mu\nu}$ itself contains such terms. This can be explicitly realized by introducing a counting parameter $\epsilon$ and replacing every term $\partial_\mu \partial_\nu g_{\rho \sigma}$ by $\epsilon \partial_\mu \partial_\nu g_{\rho \sigma}$. Doing so, we can express $A^\mu{}_{\mu}$ as a polynomial in $\epsilon$ and find, with help of the computer algebra program xAct for Mathematica~\cite{xact},
\begin{align}
A^\mu{}_\mu &= \epsilon^2 \frac {D-2}{2 (D-1)} g^{\mu\sigma} g^{\lambda \zeta} g^{\rho\omega} g^{\tau\nu}  \nonumber\\
& \bigg(
D ( -  \partial_{\zeta}\partial_{\sigma}g_{\mu\lambda} + 2  \partial_{\zeta}\partial_{\lambda}g_{\mu\sigma} ) \partial_{\nu}\partial_{\omega}g_{\rho\tau} + ( D - 1)\partial_{\omega}\partial_{\rho}g_{\mu\lambda} \partial_{\nu}\partial_{\tau}g_{\sigma \zeta}
-  D 	\partial_{\zeta}\partial_{\lambda}g_{\mu\sigma} \partial_{\nu}\partial_{\tau}g_{\rho\omega} \nonumber\\
&+ ( D - 1) \partial_{\rho}\partial_{\lambda}g_{\mu\sigma} (\partial_{\omega}\partial_{\zeta}g_{\tau\nu} - 4 \partial_{\nu}\partial_{\omega}g_{\zeta\tau} + 2 \partial_{\nu}\partial_{\tau}g_{\zeta\omega})\\
& + 2 ( D - 1) \partial_{\rho}\partial_{\sigma}g_{\mu\lambda} (\partial_{\nu}\partial_{\omega}g_{\zeta\tau} - 2 \partial_{\nu}\partial_{\tau}g_{\zeta\omega} + \partial_{\nu}\partial_{\zeta}g_{\omega\tau})\bigg) \nonumber\\
&+ \textrm{lower order terms in $\epsilon$}\,.\nonumber
\end{align}
Hence, also the untraced tensor $A_{\mu\nu}$, which is part of the truncated field equations, must contain terms of the form $\partial_\mu \partial_\nu g_{\rho \sigma} \partial_\lambda \partial_\tau g_{\zeta \omega}$.

But, any variational PDE system which is of second order must be linear in the second order derivatives acting on the fundamental dynamical variable \cite[p. 147]{Krupka-book}. Hence the truncated field equations cannot be variational and the variation of $A^\mu{}_\mu$ cannot be of second order only, but must contain higher derivatives.

\section{Necessity of densitysing in variational completion}\label{App:B}
In Section \ref{sec:varco4dgb}, we applied the variational completion algorithm to the original and to the truncated Einstein Gauss-Bonnet gravity field equations in any dimension.

An important first step in applying the algorithm was to define the densitized field equations in equation \eqref{eq:densfeq}. In the following, we are going to prove that, if the expressions $\mathcal{E}^{\mu \nu }=-\frac{1}{2}E^{\mu \nu }\sqrt{-g}$ are the Euler-Lagrange expressions of a Lagrangian $\lambda =\mathcal{L}\dd^{n}x,$ then the expressions $E^{\mu \nu }$ cannot arise as the Euler-Lagrange expressions of any Lagrangian (either coordinate-invariant or not).

Mathematically more precise, variationality is generally discussed for certain differential forms $\mathcal{E}$ on a jet bundle of a fibered
manifold, rather than for PDE's. These differential forms are called \textit{source forms} and their local coefficients $\mathcal{E}_A$ are the
left hand sides of the given PDE's. Multiplying a PDE system by a positive factor (such as $\sqrt{-g}$) will inevitably lead to a different
source form; thus, this factor does not affect the set of solutions of them PDE system, but does affect its variationality.

To fix the notation, let $(Y\overset{\pi}{\rightarrow}M,F)$ be a fiber bundle over $M$, with a local coordinate system $(x^{\mu },y^{A})$ adapted to the fibration. Sections (physically interpreted as \textit{fields})  are maps $\gamma :U\rightarrow Y$ (where $U\subset M$ is open),  locally described as $\gamma :(x^{\mu})\mapsto (y^{A}(x^{\mu })).$ On the second order jet bundle $J^{2}Y$, we denote the induced coordinates by $(x^{\mu },y^{A},y_{~\mu }^{A},y_{~\mu \nu
}^{A}).$ On the jet bundle $J^{2}Y,$ the quantities $x^{\mu },y^{A},y_{~\mu }^{A},y_{~\mu \nu }^{A}$ are interpreted
as coordinate functions (i.e., they are independent of one another); only when composed by (prolonged)\ sections, they provide  the functions
functions $(y^{A}(x^{\mu }))$ and their derivatives.

In \cite[p. 147]{Krupka-book}, it was shown that, for a second order PDE system $\mathcal{E} _{A}=\mathcal{E}_{A}(x^{\mu },y^{B},y_{~\mu }^{B},y_{~\mu \nu }^{B}),$ local variationality implies that the following \textit{Helmholtz conditions} are identically satisfied by $\mathcal{E} _{A}$:

\begin{eqnarray}
H_{AB}^{\mu \nu }(\mathcal{E} ):= &&\dfrac{\partial \mathcal{E} _{A}}{%
	\partial y_{~\mu \nu }^{B}}-\dfrac{\partial \mathcal{E} _{B}}{\partial
	y_{~\mu \nu }^{A}}=0  \label{H1} \\
H_{~AB}^{\nu }(\mathcal{E} ):= &&\dfrac{\partial \mathcal{E} _{A}}{\partial
	y_{~\nu }^{B}}+\dfrac{\partial \mathcal{E} _{B}}{\partial y_{~\nu }^{A}}%
-\dd_{\mu }\left(\dfrac{\partial \mathcal{E} _{A}}{\partial y_{~\mu \nu }^{B}}+%
\dfrac{\partial \mathcal{E} _{B}}{\partial y_{~\mu \nu }^{A}}\right)=0  \label{H2}
\\
H_{AB}(\mathcal{E} ):= &&\dfrac{\partial \mathcal{E} _{A}}{\partial y^{B}}-%
\dfrac{\partial \mathcal{E} _{B}}{\partial y^{A}}-\dfrac{1}{2}\dd_{\nu }\left(%
\dfrac{\partial \mathcal{E} _{A}}{\partial y_{~\nu }^{B}}-\dfrac{\partial
	\mathcal{E} _{B}}{\partial y_{~\nu }^{A}}\right)=0  \label{H3}
\end{eqnarray}%

Here, $\dd_{\mu }=\partial _{\mu }+y_{~\mu }^{A}\dfrac{\partial }{\partial
	y^{A}}+y_{~\mu \nu }^{A}\dfrac{\partial }{\partial y_{~\nu }^{A}}+y_{~\mu
	\nu \rho }^{A}\dfrac{\partial }{\partial y_{~\nu \rho }^{A}}$ is the total
derivative operator (of order three) acting on functions $%
f:J^{2}Y\rightarrow \mathbb{R},$ $f=f(x^{\mu },y^{B},y_{~\mu }^{B},y_{~\mu
	\nu }^{B})$. In particular, $\dd_{\mu }y^{A}=y_{~\mu }^{A}.$

\bigskip

Now, let us assume that $\mathcal{E} _{A}$ satisfies the Helmholtz conditions. Multiplying $\mathcal{E} _{A}$ by a factor $f=f(x^{\mu },y^{B})$, we obtain a new source form  $f\mathcal{E},$ with local coefficients $f\mathcal{E}_{A}.$

The first Helmholtz condition (\ref{H1}) is, indeed, not affected by the
rescaling. But, substituting $f\mathcal{E} _{A}$ instead of $\mathcal{E}
_{A} $ into (\ref{H2}) gives:%
\[
H_{~AB}^{\nu }(f\mathcal{E} ):=fH_{~AB}^{\nu }(\mathcal{E} )-(\dd_{\mu }f)\left(%
\dfrac{\partial \mathcal{E} _{A}}{\partial y_{~\mu \nu }^{B}}+\dfrac{%
	\partial \mathcal{E} _{B}}{\partial y_{~\mu \nu }^{A}}\right).
\]%
The term $fH_{~AB}^{\nu }(\mathcal{E} )$ vanishes by the variationality
assumption on $\mathcal{E} _{A};$ using (\ref{H1}) in the remaining term, we get
$\dfrac{\partial \mathcal{E} _{A}}{\partial y_{~\mu \nu }^{B}}+\dfrac{%
	\partial \mathcal{E} _{B}}{\partial y_{~\mu \nu }^{A}}=2\dfrac{\partial
	\mathcal{E} _{A}}{\partial y_{~\mu \nu }^{B}}$ and therefore:%
\begin{equation}
H_{~AB}^{\nu }(f\mathcal{E} )=-2(\dd_{\mu }f)\dfrac{\partial \mathcal{E} _{A}}{%
	\partial y_{~\mu \nu }^{B}}.  \label{H_f}
\end{equation}

In order to simplify calculations, we will contract the above equality by $y^{A}$, thus getting:
\begin{equation} 
y^{A} H_{~AB}^{\nu }(f\mathcal{E} )=-2(\dd_{\mu }f)\dfrac{\partial (y^{A}\mathcal{E} _{A})}{%
	\partial y_{~\mu \nu }^{B}}.  \label{H_f_1}
\end{equation}
A necessary (but not sufficient) condition for the variationality of the source form $f \mathcal{E}$ is then that the contracted expressions (\ref{H_f_1}) vanish.

\bigskip

Now, in order to check the above condition for the expressions $\mathcal{E}^{\mu\nu}$ studied in Section \ref{sec:varco4dgb}, we will make the following substitutions: $\mathcal{E}_{A} \to \mathcal{E}^{\mu\nu },$ $y^{A}\to g_{\mu \nu },$ $f\to \dfrac{1}{\sqrt{-g}}$.

These are functions on the jet bundle $J^{2}\Met(M)$, where $\Met(M)$ is the fiber bundle of symmetric and nondegenerate tensors of type $(0,2)$ over the spacetime manifold $M$, \cite[p. 172]{Krupka-book}. On this bundle, a system of fibered coordinate functions has the form $(x^{\mu};g_{\mu \nu };g_{\mu \nu ,\rho };g_{\mu \nu ,\rho \tau })$.

A brief direct computation, using $\dfrac{\partial g}{\partial g_{\nu \rho }}=g^{\nu \rho }g$, gives:
\[
\dd_{\mu }f= \dfrac{1}{2}(-g)^{-1/2}g^{\nu \rho }g_{\nu \rho ,\mu }= \dfrac{1}{%
	\sqrt{-g}}\Gamma _{~\mu \nu }^{\nu },
\]%
where the $\Gamma _{~\mu \nu }^{\nu }$ are \textit{formal }Christoffel
symbols, i.e., in their expressions, $x^{\mu },$ $g_{\mu \nu }$ and $g_{\mu
	\nu ,\rho }$ are all regarded as independent variables (it is only along
given \textit{sections }that we can state that $g_{\mu \nu }=g_{\mu \nu
}(x^{\rho })$). In particular, we cannot tune the coordinates $x^{\mu }$ in
such a way as to have $\Gamma _{~\mu \nu }^{\nu }=0$ even at a single point
(let alone having this equality identically satisfied).

The second factor $\dfrac{\partial (y^{A}\mathcal{E} _{A})}{\partial y_{~\mu \nu }^{B}}$ in (\ref{H_f_1}) becomes, in our case: $\dfrac{\partial(g_{\alpha \beta}\mathcal{E}^{\alpha \beta })}{\partial g_{\gamma \delta ,\mu \nu }}$. Using \eqref{eq:gbfield2} and \eqref{combi}, we find:
\begin{equation}
g_{\alpha \beta}\mathcal{E}^{\alpha \beta } = \left[\Mp^2\left(1-\dfrac{D}{2}\right)R+\Lambda_0 D- \dfrac{\alpha}{2}\mathcal{G}\right]\sqrt{-g}.
\end{equation}

The expressions \eqref{H_f_1} are then of the form:
\begin{equation}
2\Mp^2\left(1-\dfrac{D}{2}\right)\Gamma _{~\mu \tau }^{\tau }\dfrac{\partial R}{\partial g_{\gamma \delta
		,\mu \nu }}+ ... ,
\end{equation}
where the dots stand for terms which, after differentiation, will still contain curvature components; but, using the identity:\ $\dfrac{\partial R}{\partial g_{\gamma
		\delta ,\mu \nu }}=g^{\gamma \nu }g^{\delta \mu }-g^{\mu \nu }g^{\gamma
	\delta }$, the explicitly listed term above is, up to multiplication by a constant:
$(\Gamma ^{\tau \delta }{}_\tau g^{\gamma \nu }-\Gamma^{\tau \nu}{}_\tau g^{\gamma \delta })\not=0$.

Therefore, there is no chance that the full Helmholtz expressions $H^{\nu (\alpha
	\beta )(\gamma \delta )}(f\mathcal{E} )$ (which also involve nontrivial curvature terms) would identically vanish, which means that the functions $E^{\mu \nu }=-\dfrac{2}{\sqrt{%
		-g}}\mathcal{E}^{\mu \nu }$ cannot be the Euler-Lagrange expressions of any
Lagrangian (either coordinate invariant or not).

\section{Extending the Vainberg-Tonti Lagrangian}\label{App:C}
Here we prove Theorem \ref{VT-extended} of Section \ref{ssec:extVC}, which extends the definition of the Vainberg-Tonti Lagrangian to cases when the domain of definition of the functions $\mathcal{E}_{A}$  is not vertically star-shaped with center 0.

With the notations in the previous Appendix, we consider arbitrary second order PDE systems $\mathcal{E}_{A}(x^{\mu },y^{B},y_{~\mu }^{B},y_{~\mu \nu }^{B})=0$. Any such PDE system defines a source form $\varepsilon =\mathcal{E}_{A}\omega ^{A}\wedge \dd^{n}x$, where $\omega^{A}=\dd y^{A}-y_{~i}^{A}\dd x^{i}$, on some fibered chart $(V^{2},\psi^{2})$ of $J^{2}Y$. In the following, we will consider that the fibered chart domain $V^{2}$ is completely arbitrary - i.e., its image $\psi^2(V^2)$ through the coordinate homeomorphism $\psi^2$ is not necessarily vertically star-shaped.

\bigskip

\begin{lemma}
	Let $\mathcal{E}_{A} = 0$ be an arbitrary second order PDE system, and $a,b$ two arbitrary real numbers. Define, at each point in the domain of definition of $\mathcal{E}_{A}$:
	\begin{equation}
	{\mathcal{L}}_{\mathcal{E}}(x^{\mu },y^{B},y_{~\mu }^{B},y_{~\mu \nu }^{B}):=y^{A}\overset{b}{\underset{a}{\int }}%
	\mathcal{E}_{A}(x^{\mu },ty^{B},ty_{~\mu }^{B},ty_{~\mu \nu }^{B})\dd t.
	\label{extended VT Lagrangian}
	\end{equation}
	If the equations $\mathcal{E}_{A} = 0$ are locally variational and the above integrals exist and are finite, then, the Euler-Lagrange expressions \eqref{eq:eulag} associated with ${\mathcal{L}}_{\mathcal{E}}$ are:
	\begin{equation}
	\tilde{\mathcal{E}}_{B}=b \mathcal{E}_{B}(x^{\mu} b y^{A},by_{~\mu }^{A},by_{~\mu \nu }^{A})-a \mathcal{E}_{B}(x^{\mu
	},ay^{A},ay_{~\mu }^{A},ay_{~\mu \nu }^{A}).  \label{EB}
	\end{equation}
\end{lemma}

\textbf{Proof.}
Since $\mathcal{E}_{A} = 0$ are assumed to be variational, the Helmholtz
conditions (\ref{H1})-(\ref{H3}) hold. Further, let us calculate the
Euler-Lagrange expressions $\tilde{\mathcal{E}}_{B}(\mathcal{L}_{\mathcal{E}}).$ Denoting by $\chi _{t}:V^{2} \rightarrow V^{2}$ the fiber homothety $(x^{\mu },y^{B},y_{~\mu
}^{B},y_{~\mu \nu }^{B})\mapsto (x^{\mu },ty^{B},ty_{~\mu }^{B},ty_{~\mu \nu
}^{B})$ (defined for $t$ such that $\chi_{t}(V^{2}) \subset V^{2}$, where $V^{2}$ is the domain of definition of $\mathcal{E}_{A}$), we can write (\ref{extended VT Lagrangian}) in a more compact way as:
\begin{equation}
\mathcal{{L}}_{\mathcal{E}}:=y^{A}\overset{b}{\underset{a}{\int }}%
\mathcal{E} _{A}\circ \chi _{t}\dd t.
\end{equation}
Further,
\[
\dfrac{\partial \mathcal{{L}}_{\mathcal{E} }}{\partial y^{B}}=\overset{%
	b}{\underset{a}{\int }}\mathcal{E}_{B}\circ \chi _{t}\dd t+y^{A}\overset{b}{%
	\underset{a}{\int }}t\dfrac{\partial \mathcal{E}_{A}}{\partial y^{B}}\circ
\chi _{t}\dd t.
\]%
Performing integration by parts in the first term, this becomes, after a
brief computation:%
\begin{align*}
\dfrac{\partial \mathcal{{L}}_{\mathcal{E} }}{\partial y^{B}}%
&=t\mathcal{E} _{B}(x^{\mu },ty^{A},ty_{~\mu }^{A},ty_{~\mu \nu }^{A})\mid
_{a^{{}}}^{b_{{}}} \\
&+\overset{b}{\underset{a}{\int }}t[y^{A}(\dfrac{\partial \mathcal{E}_{A}}{%
	\partial y^{B}} -\dfrac{\partial \mathcal{E}_{B}}{\partial y^{A}}) \circ \chi _{t} - y_{~\mu }^{A} \dfrac{%
	\partial \mathcal{E}_{B}}{\partial y_{\mu ~}^{A}}\circ \chi _{t} -y_{~\mu \nu
}^{A} \dfrac{\partial \mathcal{E}_{B}}{\partial y_{\mu \nu ~}^{A}}\circ \chi _{t}]\dd t.
\end{align*}
The other derivatives appearing in $\tilde{\mathcal{E}}_{B}(\mathcal{{L}}_{\mathcal{E}})$ are:
\begin{align*}
\dfrac{\partial \mathcal{{L}}_{\mathcal{E} }}{\partial y_{~\mu }^{B}}
&=y^{A}\overset{b}{\underset{a}{\int }}t\dfrac{\partial \mathcal{E} _{A}}{%
	\partial y_{\mu }^{B}}\circ \chi _{t}\dd t \\
\Rightarrow\ \dd_{\mu }(\dfrac{\partial \mathcal{\tilde{L}}_{\mathcal{E} }}{\partial
	y_{~\mu }^{B}})
&=y_{~\mu }^{A}\overset{b}{\underset{a}{\int }}t\dfrac{\partial \mathcal{E} _{A}}{\partial y_{\mu }^{B}}\circ \chi _{t}\dd t
+y^{A}\overset{b}{\underset{a}{\int}}t \dd_{\mu }(\dfrac{\partial \mathcal{E} _{A}}{%
	\partial y_{\mu }^{B}}\circ \chi _{t})\dd t; \\
\dfrac{\partial \mathcal{\tilde{L}}_{\mathcal{E} }}{\partial y_{~\mu \nu}^{B}}
&=y^{A}\overset{b}{\underset{a}{\int }}t\dfrac{\partial \mathcal{E}
	_{A}}{\partial y_{\mu \nu }^{B}}\circ \chi _{t}\dd t  \\
\Rightarrow \dd_{\mu }\dd_{\nu }(\dfrac{\partial \mathcal{\tilde{L}}_{\mathcal{E} }}{%
	\partial y_{~\mu \nu }^{B}})
&=y_{~\mu \nu }^{A}\overset{b}{\underset{a}{%
		\int }}t\dfrac{\partial \mathcal{E} _{A}}{\partial y_{\mu \nu }^{B}}\circ \chi _{t} \dd t
+2y_{~\mu }^{A}\overset{b}{\underset{a}{\int }}t \dd_{\nu }(\dfrac{\partial \mathcal{E} _{A}}{\partial y_{\mu \nu }^{B}}\circ \chi _{t})\dd t
+y^{A}\overset{b}{\underset{a}{\int }}t \dd_{\mu } \dd_{\nu }(\dfrac{\partial \mathcal{E} _{A}}{\partial y_{\mu \nu }^{B}}\circ \chi _{t})\dd t.
\end{align*}

Grouping terms, we immediately find:%
\begin{align*}
\tilde{\mathcal{E}}_{B}(\mathcal{{L}}_{\mathcal{E}})
&= -\overset{b}{\underset{a}{\int }} t[y^{A}(H_{BA}\circ \chi _{t})+y_{~\mu
}^{A}(H_{BA}^{\mu }\circ \chi _{t})+y_{~\mu \nu }^{A}(H_{BA}^{\mu }\circ
\chi _{t})] \dd t,
\end{align*}
where $H_{AB}, H_{AB}^{\mu}, H_{AB}^{\mu}$ are the components of the Helmholtz form, defined in (\ref{H1})-(\ref{H3}).
By virtue of the Helmholtz conditions, these vanish identically and therefore:\
\[
\mathcal{E}_{B}(\mathcal{\tilde{L}}_{\mathcal{E} })=\left( t\mathcal{E}
_{B}(x^{\mu },ty^{A},ty_{~\mu }^{A},ty_{~\mu \nu }^{A})\right) \mid
_{a^{{}}}^{b_{{}}},
\]%
which is just (\ref{EB}).

\bigskip

The standard Vainberg-Tonti Lagrangian was determined by a similar
reasoning, choosing $b=1$ and $a=0,$ see \cite{Krupka-book}. From the
above Lemma, we immediately obtain the desired theorem:

\begin{theorem}\label{VT-extendedApp}
	Let $\mathcal{E}_{A} = 0$ be an arbitrary second order PDE system and $a \in \mathbb{R} \cup \{\pm \infty\}$ such that
	\[
	\underset{t\rightarrow a}{\lim }\left( t\mathcal{E} _{B}(x^{\mu
	},ty^{A},ty_{~\mu }^{A},ty_{~\mu \nu }^{A})\right) =0\,,
	\]%
	at all $(x^{\mu },y^{B},y_{~\mu }^{B},y_{~\mu \nu }^{B})$ in the domain of $\mathcal{E}_{A}$.	Define, at these points, the extended Vainberg-Tonti Lagrangian $\lambda = \mathcal{L} \dd^{n}x$, by the rule:
	\begin{equation}\label{eq:VTextApp}
	\mathcal{L}_{\mathcal{E} }(x^{\mu },y^{B},y_{~\mu }^{B},y_{~\mu \nu }^{B}):=y^{A}\overset{1}{\underset{a}{\int }}\mathcal{E}
	_{A}(x^{\mu },ty^{B},ty_{~\mu }^{B},ty_{~\mu \nu }^{B})\dd t\,.
	\end{equation}%
	If the above integrals exist and are finite at all points in the given domain, then:\\
	1. If the equations $\mathcal{E}_{A} = 0$ are variational, then $\lambda$ is a (locally defined) Lagrangian for these, i.e., the Euler-Lagrange expressions of \eqref{eq:VTextApp} are precisely $\mathcal{E}_{A}$:
	\begin{equation}
	\tilde{\mathcal{E}}_{A} =\mathcal{E} _{A}.
	\end{equation}

	2. If $\mathcal{E}_{A} = 0$ are not variational, then the Euler-Lagrange expressions of \eqref{eq:VTextApp} are their canonical variational completion; the correction terms are expressed in terms of the coefficients of the Helmholtz form:
	\begin{equation}
	\tilde{\mathcal{E}}_{A} = \mathcal{E} _{A} + H_A,
	\end{equation}
	where
	\begin{equation}\label{HA}
	H_A =  -\overset{b}{\underset{a}{\int }} t[y^{B}(H_{AB}\circ \chi _{t})+y_{~\mu
	}^{B}(H_{AB}^{\mu }\circ \chi _{t})+y_{~\mu \nu }^{B}(H_{AB}^{\mu }\circ
	\chi _{t})] \dd t.
	\end{equation}

\end{theorem}

\bigskip

The above result is extremely useful in the case when $\mathcal{E} _{A}$ is homogeneous of negative degree $k < -1$ in $y^{A}.$ In this case, the integral $y^{A}%
\overset{1}{\underset{a}{\int }}\mathcal{E} _{A}(x^{\mu },ty^{B},ty_{~\mu}^{B},ty_{~\mu \nu }^{B})\dd t$ diverges for $a=0$, but it can be replaced with an integral from $a=\infty $ to 1.

\bigskip

\textbf{A special case}. The case when $\mathcal{E}_{A}$ are homogeneous functions of degree $-1$ in the fiber variables, is a degenerate one. In this case, there is no integration endpoint $a$ which satisfies the hypothesis of the above Theorem. Therefore, in this case, we cannot define the Vainberg-Tonti Lagrangian \eqref{eq:VTextApp}.

\bibliographystyle{utphys}
\bibliography{varcompl}

\providecommand{\href}[2]{#2}\begingroup\raggedright\begin{thebibliography}{100}

\bibitem{Tomozawa:2011gp}
Y.~Tomozawa, ``{Quantum corrections to gravity},''
\href{http://arxiv.org/abs/1107.1424}{{\tt arXiv:1107.1424 [gr-qc]}}.

\bibitem{Cognola:2013fva}
G.~Cognola, R.~Myrzakulov, L.~Sebastiani, and S.~Zerbini, ``{Einstein gravity
  with Gauss-Bonnet entropic corrections},''
  \href{http://dx.doi.org/10.1103/PhysRevD.88.024006}{{\em Phys. Rev.} {\bf
  D88} (2013) no.~2, 024006},
\href{http://arxiv.org/abs/1304.1878}{{\tt arXiv:1304.1878 [gr-qc]}}.

\bibitem{Glavan:2019inb}
D.~Glavan and C.~Lin, ``{Einstein-Gauss-Bonnet gravity in 4-dimensional
  space-time},'' \href{http://dx.doi.org/10.1103/PhysRevLett.124.081301}{{\em
  Phys. Rev. Lett.} {\bf 124} (2020) no.~8, 081301},
\href{http://arxiv.org/abs/1905.03601}{{\tt arXiv:1905.03601 [gr-qc]}}.

\bibitem{Mann:1991qp}
R.~B. Mann, ``{Lower dimensional black holes},''
\href{http://dx.doi.org/10.1007/BF00760418}{{\em Gen. Rel. Grav.} {\bf 24}
  (1992)  433--449}.

\bibitem{Casalino:2020kbt}
A.~Casalino, A.~Colleaux, M.~Rinaldi, and S.~Vicentini, ``{Regularized Lovelock
  gravity},''
\href{http://arxiv.org/abs/2003.07068}{{\tt arXiv:2003.07068 [gr-qc]}}.

\bibitem{Lovelock:1969}
D.~Lovelock, ``{The uniqueness of the Einstein field equations in a
  four-dimensional space},'' \href{http://dx.doi.org/10.1007/BF00248156}{{\em
  Arch. Rational Mech. Anal.} {\bf 33} (1969)  54–70}.

\bibitem{Ai:2020peo}
W.-Y. Ai, ``{A note on the novel 4D Einstein-Gauss-Bonnet gravity},''
\href{http://arxiv.org/abs/2004.02858}{{\tt arXiv:2004.02858 [gr-qc]}}.

\bibitem{Gurses:2020ofy}
M.~Gurses, T.~C. Sisman, and B.~Tekin, ``{Is there a novel
  Einstein-Gauss-Bonnet theory in four dimensions?},''
\href{http://arxiv.org/abs/2004.03390}{{\tt arXiv:2004.03390 [gr-qc]}}.

\bibitem{Mahapatra:2020rds}
S.~Mahapatra, ``{A note on the total action of $4D$ Gauss-Bonnet theory},''
\href{http://arxiv.org/abs/2004.09214}{{\tt arXiv:2004.09214 [gr-qc]}}.

\bibitem{Tian:2020nzb}
S.~X. Tian and Z.-H. Zhu, ``{Comment on "Einstein-Gauss-Bonnet Gravity in
  Four-Dimensional Spacetime"},''
\href{http://arxiv.org/abs/2004.09954}{{\tt arXiv:2004.09954 [gr-qc]}}.

\bibitem{Arrechea:2020evj}
J.~Arrechea, A.~Delhom, and A.~Jiménez-Cano, ``{Yet another comment on
  four-dimensional Einstein-Gauss-Bonnet gravity},''
\href{http://arxiv.org/abs/2004.12998}{{\tt arXiv:2004.12998 [gr-qc]}}.

\bibitem{Kobayashi:2020wqy}
T.~Kobayashi, ``{Effective scalar-tensor description of regularized Lovelock
  gravity in four dimensions},''
\href{http://arxiv.org/abs/2003.12771}{{\tt arXiv:2003.12771 [gr-qc]}}.

\bibitem{Lu:2020iav}
H.~Lu and Y.~Pang, ``{Horndeski Gravity as $D\rightarrow4$ Limit of
  Gauss-Bonnet},''
\href{http://arxiv.org/abs/2003.11552}{{\tt arXiv:2003.11552 [gr-qc]}}.

\bibitem{Hennigar:2020lsl}
R.~A. Hennigar, D.~Kubiznak, R.~B. Mann, and C.~Pollack, ``{On Taking the $D\to
  4$ limit of Gauss-Bonnet Gravity: Theory and Solutions},''
\href{http://arxiv.org/abs/2004.09472}{{\tt arXiv:2004.09472 [gr-qc]}}.

\bibitem{Fernandes:2020nbq}
P.~G.~S. Fernandes, P.~Carrilho, T.~Clifton, and D.~J. Mulryne, ``{Derivation
  of Regularized Field Equations for the Einstein-Gauss-Bonnet Theory in Four
  Dimensions},''
\href{http://arxiv.org/abs/2004.08362}{{\tt arXiv:2004.08362 [gr-qc]}}.

\bibitem{Easson:2020mpq}
D.~A. Easson, T.~Manton, and A.~Svesko, ``{$D\to4$ Einstein-Gauss-Bonnet
  Gravity and Beyond},''
\href{http://arxiv.org/abs/2005.12292}{{\tt arXiv:2005.12292 [hep-th]}}.

\bibitem{Mann:1992ar}
R.~B. Mann and S.~F. Ross, ``{The $D \rightarrow 2$ limit of general
  relativity},'' \href{http://dx.doi.org/10.1088/0264-9381/10/7/015}{{\em
  Class. Quant. Grav.} {\bf 10} (1993)  1405--1408},
\href{http://arxiv.org/abs/gr-qc/9208004}{{\tt arXiv:gr-qc/9208004 [gr-qc]}}.

\bibitem{Horndeski:1974wa}
G.~W. Horndeski, ``{Second-order scalar-tensor field equations in a
  four-dimensional space},''
\href{http://dx.doi.org/10.1007/BF01807638}{{\em Int. J. Theor. Phys.} {\bf 10}
  (1974)  363--384}.

\bibitem{Lin:2020kqe}
Z.-C. Lin, K.~Yang, S.-W. Wei, Y.-Q. Wang, and Y.-X. Liu, ``{Is the
  four-dimensional novel EGB theory equivalent to its regularized counterpart
  in a cylindrically symmetric spacetime?},''
\href{http://arxiv.org/abs/2006.07913}{{\tt arXiv:2006.07913 [gr-qc]}}.

\bibitem{Aoki:2020lig}
K.~Aoki, M.~A. Gorji, and S.~Mukohyama, ``{A consistent theory of $D\rightarrow
  4$ Einstein-Gauss-Bonnet gravity},''
\href{http://arxiv.org/abs/2005.03859}{{\tt arXiv:2005.03859 [gr-qc]}}.

\bibitem{Alkac:2020zhg}
G.~Alkac and D.~O. Devecioglu, ``{Three Dimensional Modified Gravities as
  Holographic Limits of Lancsoz-Lovelock Theories},''
\href{http://arxiv.org/abs/2004.12839}{{\tt arXiv:2004.12839 [hep-th]}}.

\bibitem{Konoplya:2020bxa}
R.~A. Konoplya and A.~F. Zinhailo, ``{Quasinormal modes, stability and shadows
  of a black hole in the novel 4D Einstein-Gauss-Bonnet gravity},''
\href{http://arxiv.org/abs/2003.01188}{{\tt arXiv:2003.01188 [gr-qc]}}.

\bibitem{Guo:2020zmf}
M.~Guo and P.-C. Li, ``{The innermost stable circular orbit and shadow in the
  novel $4D$ Einstein-Gauss-Bonnet gravity},''
  \href{http://dx.doi.org/10.1140/epjc/s10052-020-8164-7}{{\em Eur. Phys. J.}
  {\bf C80} (2020) no.~6, 588},
\href{http://arxiv.org/abs/2003.02523}{{\tt arXiv:2003.02523 [gr-qc]}}.

\bibitem{Fernandes:2020rpa}
P.~G.~S. Fernandes, ``{Charged Black Holes in AdS Spaces in $4D$ Einstein
  Gauss-Bonnet Gravity},''
  \href{http://dx.doi.org/10.1016/j.physletb.2020.135468}{{\em Phys. Lett.}
  {\bf B805} (2020)  135468},
\href{http://arxiv.org/abs/2003.05491}{{\tt arXiv:2003.05491 [gr-qc]}}.

\bibitem{Konoplya:2020qqh}
R.~A. Konoplya and A.~Zhidenko, ``{Black holes in the four-dimensional
  Einstein-Lovelock gravity},''
  \href{http://dx.doi.org/10.1103/PhysRevD.101.084038}{{\em Phys. Rev.} {\bf
  D101} (2020) no.~8, 084038},
\href{http://arxiv.org/abs/2003.07788}{{\tt arXiv:2003.07788 [gr-qc]}}.

\bibitem{Wei:2020ght}
S.-W. Wei and Y.-X. Liu, ``{Testing the nature of Gauss-Bonnet gravity by
  four-dimensional rotating black hole shadow},''
\href{http://arxiv.org/abs/2003.07769}{{\tt arXiv:2003.07769 [gr-qc]}}.

\bibitem{Kumar:2020owy}
R.~Kumar and S.~G. Ghosh, ``{Rotating black holes in the novel $4D$
  Einstein-Gauss-Bonnet gravity},''
\href{http://arxiv.org/abs/2003.08927}{{\tt arXiv:2003.08927 [gr-qc]}}.

\bibitem{Hegde:2020xlv}
K.~Hegde, A.~Naveena~Kumara, C.~L.~A. Rizwan, A.~K. M., and M.~S. Ali,
  ``{Thermodynamics, Phase Transition and Joule Thomson Expansion of novel 4-D
  Gauss Bonnet AdS Black Hole},''
\href{http://arxiv.org/abs/2003.08778}{{\tt arXiv:2003.08778 [gr-qc]}}.

\bibitem{Ghosh:2020vpc}
S.~G. Ghosh and S.~D. Maharaj, ``{Radiating black holes in the novel 4D
  Einstein-Gauss-Bonnet gravity},''
\href{http://arxiv.org/abs/2003.09841}{{\tt arXiv:2003.09841 [gr-qc]}}.

\bibitem{Zhang:2020qew}
Y.-P. Zhang, S.-W. Wei, and Y.-X. Liu, ``{Spinning test particle in
  four-dimensional Einstein-Gauss-Bonnet Black Hole},''
\href{http://arxiv.org/abs/2003.10960}{{\tt arXiv:2003.10960 [gr-qc]}}.

\bibitem{Singh:2020xju}
D.~V. Singh and S.~Siwach, ``{Thermodynamics and P-v criticality of Bardeen-AdS
  Black Hole in 4-D Einstein-Gauss-Bonnet Gravity},''
\href{http://arxiv.org/abs/2003.11754}{{\tt arXiv:2003.11754 [gr-qc]}}.

\bibitem{Konoplya:2020ibi}
R.~A. Konoplya and A.~Zhidenko, ``{BTZ black holes with higher curvature
  corrections in the 3D Einstein-Lovelock theory},''
\href{http://arxiv.org/abs/2003.12171}{{\tt arXiv:2003.12171 [gr-qc]}}.

\bibitem{Ghosh:2020syx}
S.~G. Ghosh and R.~Kumar, ``{Generating black holes in the novel $4D$
  Einstein-Gauss-Bonnet gravity},''
\href{http://arxiv.org/abs/2003.12291}{{\tt arXiv:2003.12291 [gr-qc]}}.

\bibitem{Konoplya:2020juj}
R.~A. Konoplya and A.~Zhidenko, ``{(In)stability of black holes in the 4D
  Einstein-Gauss-Bonnet and Einstein-Lovelock gravities},''
\href{http://arxiv.org/abs/2003.12492}{{\tt arXiv:2003.12492 [gr-qc]}}.

\bibitem{Zhang:2020qam}
C.-Y. Zhang, P.-C. Li, and M.~Guo, ``{Greybody factor and power spectra of the
  Hawking radiation in the novel $4D$ Einstein-Gauss-Bonnet de-Sitter
  gravity},''
\href{http://arxiv.org/abs/2003.13068}{{\tt arXiv:2003.13068 [hep-th]}}.

\bibitem{HosseiniMansoori:2020yfj}
S.~A. Hosseini~Mansoori, ``{Thermodynamic geometry of the novel 4-D Gauss
  Bonnet AdS Black Hole},''
\href{http://arxiv.org/abs/2003.13382}{{\tt arXiv:2003.13382 [gr-qc]}}.

\bibitem{Kumar:2020uyz}
A.~Kumar and R.~Kumar, ``{Bardeen black holes in the novel $4D$
  Einstein-Gauss-Bonnet gravity},''
\href{http://arxiv.org/abs/2003.13104}{{\tt arXiv:2003.13104 [gr-qc]}}.

\bibitem{Roy:2020dyy}
R.~Roy and S.~Chakrabarti, ``{A study on black hole shadows in asymptotically
  de Sitter spacetimes},''
\href{http://arxiv.org/abs/2003.14107}{{\tt arXiv:2003.14107 [gr-qc]}}.

\bibitem{Singh:2020nwo}
D.~V. Singh, S.~G. Ghosh, and S.~D. Maharaj, ``{Clouds of string in the novel
  $4D$ Einstein-Gauss-Bonnet black holes},''
\href{http://arxiv.org/abs/2003.14136}{{\tt arXiv:2003.14136 [gr-qc]}}.

\bibitem{Wei:2020poh}
S.-W. Wei and Y.-X. Liu, ``{Extended thermodynamics and microstructures of
  four-dimensional charged Gauss-Bonnet black hole in AdS space},''
  \href{http://dx.doi.org/10.1103/PhysRevD.101.104018}{{\em Phys. Rev.} {\bf
  D101} (2020) no.~10, 104018},
\href{http://arxiv.org/abs/2003.14275}{{\tt arXiv:2003.14275 [gr-qc]}}.

\bibitem{Churilova:2020aca}
M.~S. Churilova, ``{Quasinormal modes of the Dirac field in the novel 4D
  Einstein-Gauss-Bonnet gravity},''
\href{http://arxiv.org/abs/2004.00513}{{\tt arXiv:2004.00513 [gr-qc]}}.

\bibitem{Kumar:2020xvu}
A.~Kumar and S.~G. Ghosh, ``{Hayward black holes in the novel $4D$
  Einstein-Gauss-Bonnet gravity},''
\href{http://arxiv.org/abs/2004.01131}{{\tt arXiv:2004.01131 [gr-qc]}}.

\bibitem{Islam:2020xmy}
S.~U. Islam, R.~Kumar, and S.~G. Ghosh, ``{Gravitational lensing by black holes
  in $4D$ Einstein-Gauss-Bonnet gravity},''
\href{http://arxiv.org/abs/2004.01038}{{\tt arXiv:2004.01038 [gr-qc]}}.

\bibitem{Mishra:2020gce}
A.~K. Mishra, ``{Quasinormal modes and Strong Cosmic Censorship in the novel 4D
  Einstein-Gauss-Bonnet gravity},''
\href{http://arxiv.org/abs/2004.01243}{{\tt arXiv:2004.01243 [gr-qc]}}.

\bibitem{Liu:2020vkh}
C.~Liu, T.~Zhu, and Q.~Wu, ``{Thin Accretion Disk around a four-dimensional
  Einstein-Gauss-Bonnet Black Hole},''
\href{http://arxiv.org/abs/2004.01662}{{\tt arXiv:2004.01662 [gr-qc]}}.

\bibitem{Konoplya:2020cbv}
R.~A. Konoplya and A.~F. Zinhailo, ``{Grey-body factors and Hawking radiation
  of black holes in $4D$ Einstein-Gauss-Bonnet gravity},''
\href{http://arxiv.org/abs/2004.02248}{{\tt arXiv:2004.02248 [gr-qc]}}.

\bibitem{Jin:2020emq}
X.-H. Jin, Y.-X. Gao, and D.-J. Liu, ``{Strong gravitational lensing of a
  4-dimensional Einstein-Gauss-Bonnet black hole in homogeneous plasma},''
\href{http://arxiv.org/abs/2004.02261}{{\tt arXiv:2004.02261 [gr-qc]}}.

\bibitem{Heydari-Fard:2020sib}
M.~Heydari-Fard, M.~Heydari-Fard, and H.~R. Sepangi, ``{Bending of light in
  novel 4$D$ Gauss-Bonnet-de Sitter black holes by Rindler-Ishak method},''
\href{http://arxiv.org/abs/2004.02140}{{\tt arXiv:2004.02140 [gr-qc]}}.

\bibitem{Zhang:2020sjh}
C.-Y. Zhang, S.-J. Zhang, P.-C. Li, and M.~Guo, ``{Superradiance and stability
  of the novel 4D charged Einstein-Gauss-Bonnet black hole},''
\href{http://arxiv.org/abs/2004.03141}{{\tt arXiv:2004.03141 [gr-qc]}}.

\bibitem{EslamPanah:2020hoj}
B.~Eslam~Panah and K.~Jafarzade, ``{4D Einstein-Gauss-Bonnet AdS Black Holes as
  Heat Engine},''
\href{http://arxiv.org/abs/2004.04058}{{\tt arXiv:2004.04058 [hep-th]}}.

\bibitem{NaveenaKumara:2020rmi}
A.~Naveena~Kumara, C.~L.~A. Rizwan, K.~Hegde, M.~S. Ali, and A.~K. M,
  ``{Rotating 4D Gauss-Bonnet black hole as particle accelerator},''
\href{http://arxiv.org/abs/2004.04521}{{\tt arXiv:2004.04521 [gr-qc]}}.

\bibitem{Aragon:2020qdc}
A.~Aragón, R.~Bécar, P.~A. González, and Y.~Vásquez, ``{Perturbative and
  nonperturbative quasinormal modes of 4D Einstein-Gauss-Bonnet black holes},''
\href{http://arxiv.org/abs/2004.05632}{{\tt arXiv:2004.05632 [gr-qc]}}.

\bibitem{Yang:2020czk}
S.-J. Yang, J.-J. Wan, J.~Chen, J.~Yang, and Y.-Q. Wang, ``{Weak cosmic
  censorship conjecture for the novel $4D$ charged Einstein-Gauss-Bonnet black
  hole with test scalar field and particle},''
\href{http://arxiv.org/abs/2004.07934}{{\tt arXiv:2004.07934 [gr-qc]}}.

\bibitem{Cuyubamba:2020moe}
M.~A. Cuyubamba, ``{Stability of asymptotically de Sitter and anti-de Sitter
  black holes in $4D$ regularized Einstein-Gauss-Bonnet theory},''
\href{http://arxiv.org/abs/2004.09025}{{\tt arXiv:2004.09025 [gr-qc]}}.

\bibitem{Ying:2020bch}
S.~Ying, ``{Thermodynamics and Weak Cosmic Censorship Conjecture of 4D
  Gauss-Bonnet-Maxwell Black Holes via Charged Particle Absorption},''
\href{http://arxiv.org/abs/2004.09480}{{\tt arXiv:2004.09480 [gr-qc]}}.

\bibitem{Rayimbaev:2020lmz}
J.~Rayimbaev, A.~Abdujabbarov, B.~Turimov, and F.~Atamurotov, ``Dynamics of
  magnetized particles around 4-d einstein gauss–bonnet black hole,''
  \href{http://dx.doi.org/https://doi.org/10.1016/j.dark.2020.100715}{{\em
  Physics of the Dark Universe} {\bf 30} (2020)  100715},
  \href{http://arxiv.org/abs/2004.10031}{{\tt arXiv:2004.10031 [gr-qc]}}.
\url{http://www.sciencedirect.com/science/article/pii/S2212686420304283}.

\bibitem{Liu:2020evp}
P.~Liu, C.~Niu, and C.-Y. Zhang, ``{Instability of the novel 4D charged
  Einstein-Gauss-Bonnet de-Sitter black hole},''
\href{http://arxiv.org/abs/2004.10620}{{\tt arXiv:2004.10620 [gr-qc]}}.

\bibitem{Zeng:2020dco}
X.-X. Zeng, H.-Q. Zhang, and H.~Zhang, ``{Shadows and photon spheres with
  spherical accretions in the four-dimensional Gauss-Bonnet black hole},''
\href{http://arxiv.org/abs/2004.12074}{{\tt arXiv:2004.12074 [gr-qc]}}.

\bibitem{Ge:2020tid}
X.-H. Ge and S.-J. Sin, ``{Causality of black holes in 4-dimensional
  Einstein-Gauss-Bonnet-Maxwell theory},''
\href{http://arxiv.org/abs/2004.12191}{{\tt arXiv:2004.12191 [hep-th]}}.

\bibitem{Hennigar:2020fkv}
R.~A. Hennigar, D.~Kubiznak, R.~B. Mann, and C.~Pollack, ``{Lower-dimensional
  Gauss-Bonnet Gravity and BTZ Black Holes},''
\href{http://arxiv.org/abs/2004.12995}{{\tt arXiv:2004.12995 [gr-qc]}}.

\bibitem{Kumar:2020sag}
R.~Kumar, S.~U. Islam, and S.~G. Ghosh, ``{Gravitational lensing by Charged
  black hole in regularized $4D$ Einstein-Gauss-Bonnet gravity},''
\href{http://arxiv.org/abs/2004.12970}{{\tt arXiv:2004.12970 [gr-qc]}}.

\bibitem{Ghosh:2020cob}
S.~G. Ghosh and S.~D. Maharaj, ``{Noncommutative inspired black holes in
  regularised 4D Einstein-Gauss-Bonnet theory},''
\href{http://arxiv.org/abs/2004.13519}{{\tt arXiv:2004.13519 [gr-qc]}}.

\bibitem{Churilova:2020mif}
M.~S. Churilova, ``{Quasinormal modes of the test fields in the novel 4D
  Einstein-Gauss-Bonnet-de Sitter gravity},''
\href{http://arxiv.org/abs/2004.14172}{{\tt arXiv:2004.14172 [gr-qc]}}.

\bibitem{Yang:2020jno}
K.~Yang, B.-M. Gu, S.-W. Wei, and Y.-X. Liu, ``{Born-Infeld Black Holes in
  novel 4D Einstein-Gauss-Bonnet gravity},''
\href{http://arxiv.org/abs/2004.14468}{{\tt arXiv:2004.14468 [gr-qc]}}.

\bibitem{Devi:2020uac}
S.~Devi, R.~Roy, and S.~Chakrabarti, ``{Quasinormal modes and greybody factors
  of the novel four dimensional Gauss-Bonnet black holes in asymptotically de
  Sitter space time: Scalar, Electromagnetic and Dirac perturbations},''
\href{http://arxiv.org/abs/2004.14935}{{\tt arXiv:2004.14935 [gr-qc]}}.

\bibitem{Jusufi:2020qyw}
K.~Jusufi, ``{Nonlinear magnetically charged black holes in 4D
  Einstein-Gauss-Bonnet gravity},''
\href{http://arxiv.org/abs/2005.00360}{{\tt arXiv:2005.00360 [gr-qc]}}.

\bibitem{Konoplya:2020der}
R.~A. Konoplya and A.~Zhidenko, ``{4D Einstein-Lovelock black holes: Hierarchy
  of orders in curvature},''
\href{http://arxiv.org/abs/2005.02225}{{\tt arXiv:2005.02225 [gr-qc]}}.

\bibitem{Liu:2020lwc}
P.~Liu, C.~Niu, and C.-Y. Zhang, ``{Instability of the Novel 4D Charged
  Einstein-Gauss-Bonnet Anti de-Sitter Black Hole},''
\href{http://arxiv.org/abs/2005.01507}{{\tt arXiv:2005.01507 [gr-qc]}}.

\bibitem{Qiao:2020hkx}
X.~Qiao, L.~OuYang, D.~Wang, Q.~Pan, and J.~Jing, ``{Holographic
  superconductors in 4D Einstein-Gauss-Bonnet gravity},''
\href{http://arxiv.org/abs/2005.01007}{{\tt arXiv:2005.01007 [hep-th]}}.

\bibitem{Dadhich:2020ukj}
N.~Dadhich, ``{On causal structure of $4D$-Einstein-Gauss-Bonnet black hole},''
\href{http://arxiv.org/abs/2005.05757}{{\tt arXiv:2005.05757 [gr-qc]}}.

\bibitem{Shaymatov:2020yte}
S.~Shaymatov, J.~Vrba, D.~Malafarina, B.~Ahmedov, and Z.~Stuchlík, ``{Charged
  particle and epicyclic motions around $4D$ Einstein-Gauss-Bonnet black hole
  immersed in an external magnetic field},''
  \href{http://dx.doi.org/10.1016/j.dark.2020.100648}{{\em Phys. Dark Univ.}
  {\bf 30} (2020)  100648},
\href{http://arxiv.org/abs/2005.12410}{{\tt arXiv:2005.12410 [gr-qc]}}.

\bibitem{Hennigar:2020zif}
R.~A. Hennigar, D.~Kubiznak, and R.~B. Mann, ``{Rotating Gauss-Bonnet BTZ Black
  Holes},''
\href{http://arxiv.org/abs/2005.13732}{{\tt arXiv:2005.13732 [gr-qc]}}.

\bibitem{Singh:2020mty}
D.~V. Singh, R.~Kumar, S.~G. Ghosh, and S.~D. Maharaj, ``{Phase transition of
  AdS black holes in 4D EGB gravity coupled to nonlinear electrodynamics},''
\href{http://arxiv.org/abs/2006.00594}{{\tt arXiv:2006.00594 [gr-qc]}}.

\bibitem{Malafarina:2020pvl}
D.~Malafarina, B.~Toshmatov, and N.~Dadhich, ``{Dust collapse in 4D
  Einstein-Gauss-Bonnet gravity},''
  \href{http://dx.doi.org/10.1016/j.dark.2020.100598}{{\em Phys. Dark Univ.}
  {\bf 30} (2020)  100598},
\href{http://arxiv.org/abs/2004.07089}{{\tt arXiv:2004.07089 [gr-qc]}}.

\bibitem{Jusufi:2020yus}
K.~Jusufi, A.~Banerjee, and S.~G. Ghosh, ``{Wormholes in 4D
  Einstein-Gauss-Bonnet Gravity},''
\href{http://arxiv.org/abs/2004.10750}{{\tt arXiv:2004.10750 [gr-qc]}}.

\bibitem{Liu:2020yhu}
P.~Liu, C.~Niu, X.~Wang, and C.-Y. Zhang, ``{Traversable Thin-shell Wormhole in
  the Novel 4D Einstein-Gauss-Bonnet Theory},''
\href{http://arxiv.org/abs/2004.14267}{{\tt arXiv:2004.14267 [gr-qc]}}.

\bibitem{Chew:2020lkj}
X.~Y. Chew, G.~Tumurtushaa, and D.-h. Yeom, ``{Euclidean wormholes in
  Gauss-Bonnet-dilaton gravity},''
\href{http://arxiv.org/abs/2006.04344}{{\tt arXiv:2006.04344 [gr-qc]}}.

\bibitem{Doneva:2020ped}
D.~D. Doneva and S.~S. Yazadjiev, ``{Relativistic stars in 4D
  Einstein-Gauss-Bonnet gravity},''
\href{http://arxiv.org/abs/2003.10284}{{\tt arXiv:2003.10284 [gr-qc]}}.

\bibitem{Banerjee:2020stc}
A.~Banerjee and K.~N. Singh, ``{Color flavor locked strange stars in 4D
  Einstein-Gauss-Bonnet gravity},''
\href{http://arxiv.org/abs/2005.04028}{{\tt arXiv:2005.04028 [gr-qc]}}.

\bibitem{Banerjee:2020yhu}
A.~Banerjee, T.~Tangphati, and P.~Channuie, ``{Strange Quark Stars in 4D
  Einstein-Gauss-Bonnet Gravity},''
\href{http://arxiv.org/abs/2006.00479}{{\tt arXiv:2006.00479 [gr-qc]}}.

\bibitem{Li:2020tlo}
S.-L. Li, P.~Wu, and H.~Yu, ``{Stability of the Einstein Static Universe in $4
  D$ Gauss-Bonnet Gravity},''
\href{http://arxiv.org/abs/2004.02080}{{\tt arXiv:2004.02080 [gr-qc]}}.

\bibitem{Ma:2020ufk}
L.~Ma and H.~Lu, ``{Vacua and Exact Solutions in Lower-$D$ Limits of EGB},''
\href{http://arxiv.org/abs/2004.14738}{{\tt arXiv:2004.14738 [gr-qc]}}.

\bibitem{Samart:2020sxj}
D.~Samart and P.~Channuie, ``{Generalized gravitational phase transition in
  novel 4D Einstein-Gauss-Bonnet gravity},''
\href{http://arxiv.org/abs/2005.02826}{{\tt arXiv:2005.02826 [gr-qc]}}.

\bibitem{Narain:2020qhh}
G.~Narain and H.-Q. Zhang, ``{Cosmic evolution in novel-Gauss Bonnet
  Gravity},''
\href{http://arxiv.org/abs/2005.05183}{{\tt arXiv:2005.05183 [gr-qc]}}.

\bibitem{Aoki:2020iwm}
K.~Aoki, M.~A. Gorji, and S.~Mukohyama, ``{Cosmology and gravitational waves in
  consistent $D\to 4$ Einstein-Gauss-Bonnet gravity},''
\href{http://arxiv.org/abs/2005.08428}{{\tt arXiv:2005.08428 [gr-qc]}}.

\bibitem{MohseniSadjadi:2020jmc}
H.~Mohseni~Sadjadi, ``{On cosmic acceleration in four dimensional
  Einstein-Gauss-Bonnet gravity},''
\href{http://arxiv.org/abs/2005.10024}{{\tt arXiv:2005.10024 [gr-qc]}}.

\bibitem{Clifton:2020xhc}
T.~Clifton, P.~Carrilho, P.~G.~S. Fernandes, and D.~J. Mulryne,
  ``{Observational Constraints on the Regularized 4D Einstein-Gauss-Bonnet
  Theory of Gravity},''
\href{http://arxiv.org/abs/2006.15017}{{\tt arXiv:2006.15017 [gr-qc]}}.

\bibitem{Casalino:2020pyv}
A.~Casalino and L.~Sebastiani, ``{Perturbations in Regularized Lovelock
  Gravity},''
\href{http://arxiv.org/abs/2004.10229}{{\tt arXiv:2004.10229 [gr-qc]}}.

\bibitem{Haghani:2020ynl}
Z.~Haghani, ``{Growth of matter density perturbations in 4D
  Einstein-Gauss-Bonnet gravity},''
\href{http://arxiv.org/abs/2005.01636}{{\tt arXiv:2005.01636 [gr-qc]}}.

\bibitem{Feng:2020duo}
J.-X. Feng, B.-M. Gu, and F.-W. Shu, ``{Theoretical and observational
  constraints on regularized 4$D$ Einstein-Gauss-Bonnet gravity},''
\href{http://arxiv.org/abs/2006.16751}{{\tt arXiv:2006.16751 [gr-qc]}}.

\bibitem{Lu:2020mjp}
H.~Lu and P.~Mao, ``{Asymptotic structure of Einstein-Gauss-Bonnet theory in
  lower dimensions},''
\href{http://arxiv.org/abs/2004.14400}{{\tt arXiv:2004.14400 [hep-th]}}.

\bibitem{Shu:2020cjw}
F.-W. Shu, ``{Vacua in novel 4D Einstein-Gauss-Bonnet Gravity: pathology and
  instability?},''
\href{http://arxiv.org/abs/2004.09339}{{\tt arXiv:2004.09339 [gr-qc]}}.

\bibitem{Bonifacio:2020vbk}
J.~Bonifacio, K.~Hinterbichler, and L.~A. Johnson, ``{Amplitudes and 4D
  Gauss-Bonnet Theory},''
\href{http://arxiv.org/abs/2004.10716}{{\tt arXiv:2004.10716 [hep-th]}}.

\bibitem{Narain:2020tsw}
G.~Narain and H.-Q. Zhang, ``{Lorentzian quantum cosmology in novel
  Gauss-Bonnet gravity from Picard-Lefschetz methods},''
\href{http://arxiv.org/abs/2006.02298}{{\tt arXiv:2006.02298 [gr-qc]}}.

\bibitem{Voicu:2015dxa}
N.~Voicu and D.~Krupka, ``{Canonical variational completion of differential
  equations},'' \href{http://dx.doi.org/10.1063/1.4918789}{{\em J. Math. Phys.}
  {\bf 56} (2015) no.~4, 043507},
\href{http://arxiv.org/abs/1406.6646}{{\tt arXiv:1406.6646 [math-ph]}}.

\bibitem{Krupka-book}
D.~Krupka, {\em Introduction to Global Variational Geometry}.
\newblock Springer, 2015.

\bibitem{Hohmann:2019sni}
M.~Hohmann, C.~Pfeifer, and N.~Voicu, ``{Relativistic kinetic gases as direct
  sources of gravity},''
  \href{http://dx.doi.org/10.1103/PhysRevD.101.024062}{{\em Phys. Rev. D} {\bf
  101} (2020) no.~2, 024062}, \href{http://arxiv.org/abs/1910.14044}{{\tt
  arXiv:1910.14044 [gr-qc]}}.

\bibitem{xact}
J.~M. Martín-García, {\em xAct: Efficient tensor computer algebra for
  Mathematica}.
\newblock 2002 - 2020.
\newblock \url{http://xact.es/}.

\bibitem{xtras}
T.~Nutma, ``{xTras : A field-theory inspired xAct package for mathematica},''
  \href{http://dx.doi.org/10.1016/j.cpc.2014.02.006}{{\em Comput. Phys.
  Commun.} {\bf 185} (2014)  1719--1738},
\href{http://arxiv.org/abs/1308.3493}{{\tt arXiv:1308.3493 [cs.SC]}}.

\bibitem{Nojiri:2018ouv}
S.~Nojiri, S.~Odintsov, and V.~Oikonomou, ``{Ghost-free Gauss-Bonnet Theories
  of Gravity},'' \href{http://dx.doi.org/10.1103/PhysRevD.99.044050}{{\em Phys.
  Rev. D} {\bf 99} (2019) no.~4, 044050},
  \href{http://arxiv.org/abs/1811.07790}{{\tt arXiv:1811.07790 [gr-qc]}}.

\bibitem{Nojiri:2005vv}
S.~Nojiri, S.~D. Odintsov, and M.~Sasaki, ``{Gauss-Bonnet dark energy},''
  \href{http://dx.doi.org/10.1103/PhysRevD.71.123509}{{\em Phys. Rev. D} {\bf
  71} (2005)  123509}, \href{http://arxiv.org/abs/hep-th/0504052}{{\tt
  arXiv:hep-th/0504052}}.

\bibitem{DeLaurentis:2015fea}
M.~De~Laurentis, M.~Paolella, and S.~Capozziello, ``{Cosmological inflation in
  $F(R,\mathcal{G})$ gravity},''
  \href{http://dx.doi.org/10.1103/PhysRevD.91.083531}{{\em Phys. Rev. D} {\bf
  91} (2015) no.~8, 083531}, \href{http://arxiv.org/abs/1503.04659}{{\tt
  arXiv:1503.04659 [gr-qc]}}.

\bibitem{Capozziello:2014ioa}
S.~Capozziello, M.~De~Laurentis, and S.~D. Odintsov, ``{Noether Symmetry
  Approach in Gauss-Bonnet Cosmology},''
  \href{http://dx.doi.org/10.1142/S0217732314501648}{{\em Mod. Phys. Lett. A}
  {\bf 29} (2014) no.~30, 1450164}, \href{http://arxiv.org/abs/1406.5652}{{\tt
  arXiv:1406.5652 [gr-qc]}}.

\bibitem{Bajardi:2020osh}
F.~Bajardi and S.~Capozziello, ``{$f(\mathcal {G})$ Noether cosmology},''
  \href{http://dx.doi.org/10.1140/epjc/s10052-020-8258-2}{{\em Eur. Phys. J.}
  {\bf C80} (2020) no.~8, 704},
\href{http://arxiv.org/abs/2005.08313}{{\tt arXiv:2005.08313 [gr-qc]}}.

\end{thebibliography}\endgroup
\end{document}